\documentclass[12pt,eqsecnum]{article}
\usepackage[dvips]{graphicx}
\usepackage{amssymb}
\usepackage{amsmath}
\usepackage{epstopdf}
\makeatletter

\@addtoreset{equation}{section}
\makeatletter

\newcommand{\D}{{\rm d}}

\newcommand{\dalm}{\kern1pt\vbox{\hrule height 0.9pt\hbox{\vrule width
0.9pt\hskip 2.5pt\vbox{\vskip 5.5pt}\hskip 3pt\vrule width 0.3pt}\hrule height
0.3pt}\kern1pt}

\def\b2hat{ {\hat b}_2 }

\def\be {\begin{equation}}
\def\ee  {\end{equation}}
\def\bea {\begin{eqnarray}}
\def\eea {\end{eqnarray}}

\begin{document}

\begin{titlepage}
\vfill
\begin{flushright}
\today
\end{flushright}

\vfill
\begin{center}
\baselineskip=16pt
{\Large\bf 
Unitary evolution of the quantum universe with a Brown-Kucha\v{r} dust\\
}
\vskip 0.5cm
{\large {\sl }}
\vskip 10.mm
{\bf Hideki Maeda} \\

\vskip 1cm
{
	Department of Electronics and Information Engineering, Hokkai-Gakuen University, Sapporo 062-8605, Japan\\
	\texttt{h-maeda-at-hgu.jp}

     }
\vspace{6pt}
\today
\end{center}
\vskip 0.2in
\par
\begin{center}
{\bf Abstract}
\end{center}
\begin{quote}
We study the time evolution of a wave function for the spatially flat Friedmann-Lema{\^ i}tre-Robertson-Walker universe governed by the Wheeler-DeWitt equation in both analytical and numerical methods. 
We consider a Brown-Kucha\v{r} dust as a matter field in order to introduce a "clock" in quantum cosmology and adopt the Laplace-Beltrami operator-ordering. 
The Hamiltonian operator admits an infinite number of self-adjoint extensions corresponding to a one-parameter family of boundary conditions at the origin in the minisuperspace.
For any value of the extension parameter in the boundary condition, the evolution of a wave function is unitary and the classical initial singularity is avoided and replaced by the big bounce in the quantum system.
Exact wave functions show that the expectation value of the spatial volume of the universe obeys the classical time evolution  in the late time but its variance diverges.
\vfill
\vskip 2.mm
\end{quote}
\end{titlepage}




\tableofcontents

\newpage

\section{Introduction}
Now the Inflation-Big-Bang scenario is undoubtedly a big paradigm in cosmology, supported by the rapid development of observation technologies in the modern era~\cite{BICEP2,planck}.
Nevertheless, the singularity theorems in general relativity assert that there appears the initial singularity quite generically and classical physics breaks down at the very early stage of the universe~\cite{singularities}.
The resolution of this initial-singularity problem requires the quantum description of the universe, that is, quantum cosmology.
(See~\cite{qc-review} for a review.)

The oldest approach to quantum gravity is the canonical approach pioneered by DeWitt~\cite{dewitt1967}, based on the Arnowitt-Deser-Misner (ADM) formalism of the Einstein equations~\cite{adm}.
Using the following ADM metric;
\begin{align}
\D s^2=-N^2\D t^2+h_{ij}(N^i\D t+\D x^i)(N^j\D t+\D x^j),
\end{align}
where $N$, $N^i$, and $h_{ij}$ are functions of $t(\equiv x^0)$ and $x^i~(i=1,2,3)$, we can write down the Einstein equations in the form of a constrained dynamical system.
In this ADM formulation, the Einstein-Hilbert action $S_{\rm G}$ is written as
\begin{align}
S_{\rm G}=\int \D t \D ^3x\left(p^{ij}\frac{\partial h_{ij}}{\partial t}-NH-N_iH^i\right)+\mbox{(total derivative)}.
\end{align}
Here $p^{ij}$ is the momentum conjugate of $h_{ij}$, which is the spatial metric  on a hypersurface $\Sigma$ with constant $t$.
$N$ and $N^i$ act as the Lagrange multipliers corresponding to the constraints $H=0$ and $H^i=0$, respectively, where the super-momentum $H^i$ and super-Hamiltonian $H$ are functionals of $h_{ij}$ and $p^{ij}$.
The momentum constraints $H^i=0$ generate spatial diffeomorphisms, while the Hamiltonian constraint $H=0$ generates time reparametrizations.

One of the possible quantizations of such a constrained dynamical system is the Dirac quantization~\cite{Dirac} in which the constraint equations become operators acting on the wave function(al) of the spacetime $\Psi[h_{ij}]$.
The resulting quantum versions of the momentum constraints ${\hat H}^i\Psi=0$ are formally satisfied by considering the DeWitt superspace in which $h_{ij}$ take values in the quotient space of $h_{ij}$ under the action of the group of spatial diffeomorphisms. 
Finally, the remaining basic equation in canonical quantum gravity is the Hamiltonian constraint ${\hat H}\Psi=0$, called the Wheeler-DeWitt equation.
Here there is an ambiguity of the operator-ordering when we replace the momentum conjugates $p^{ij}$ by operators ${\hat p}^{ij}:=-i\hbar \delta/\delta h_{ij}$.
In the present paper, we adopt the natural Laplace-Beltrami operator-ordering in the DeWitt superspace, originally proposed by Christodoulakis and Zanelli~\cite{zanelli1986}.

At a glance, the Wheeler-DeWitt equation has the form of the stationary Schr\"odinger equation with zero energy.
However, since the metric in the superspace called the DeWitt supermetric has a Lorentzian $(-,+,+,+,+,+)$ signature at each point $x^i\in \Sigma$, the Wheeler-DeWitt equation actually has the form of the Klein-Gordon equation in the superspace~\cite{Kiefer}.
(See also~\cite{kuchar} for a review of the problem of time in quantum gravity.)
This property of the Wheeler-DeWitt equation causes a problem of the conserved inner product.
To avoid this problem, Brown and Kucha\v{r} introduced "time" (or a "clock") by matter fields which provide a privileged dynamical reference frame~\cite{bk1995}.
More precisely, they introduced the so-called Brown-Kucha\v{r} dust which consists of a set of non-canonical scalar fields equivalent to a single timelike dust fluid and cast the Wheeler-DeWitt equation into the form of the time-dependent Schr\"odinger equation.
(See also~\cite{hp2011}.)

In quantum cosmology, one considers only spatially homogeneous cosmological spacetimes to be quantized.
In this minisuperspace approach, the metric functions depend only on the time coordinate, so that the resulting theory is not a quantum field theory but just quantum mechanics.
For example, in the case of the Friedmann-Lema{\^ i}tre-Robertson-Walker (FLRW) minisuperspace, the effective gravitational action is
\begin{align}
S_{\rm G}\propto \int \D t\left(p\frac{\D a}{\D t}-NH\right)+\mbox{(total derivative)},
\end{align}
where $a=a(t)$ is the scale factor of the universe and $p=p(t)$ is its conjugate.
The initial singularity at $a(t)=0$ in the classical theory is cured in quantum cosmology if the corresponding quantum mechanics is well-defined.

In~\cite{ak}, Amemiya and Koike studied the spatially flat FLRW quantum cosmology with a Brown-Kucha\v{r} dust in the presence of a cosmological constant under three conceivable operator-orderings which are different from the Laplace-Beltrami one.
The resulting Wheeler-DeWitt equation has the form of the time-dependent Schr\"odinger equation on the half-line and then the quantum cosmology is well-defined if the Hamiltonian operator ${\hat H}$ acting on the wave function of the universe $\Psi=\Psi[a]$ is self-adjoint or admit self-adjoint extensions\footnote{The earliest studies of the self-adjointness of the Hamiltonian operator in the self-gravitating system were in the context of gravitational collapse of a timelike dust shell~\cite{hajicek1992,hkk1992}.}.
It was proved that, under all the operator-orderings, the Hamiltonian operator admits an infinite number of self-adjoint extensions corresponding to a one-parameter family of boundary conditions for the wave function at $a=0$\footnote{The situation is similar in the system of a quantum harmonic oscillator on the half-line, in which the energy spectrum depends sharply on the value of the extension parameter. (See Appendix B in~\cite{km2014}.)}.
Amemiya and Koike finally showed that the classical initial singularity is replaced by a big bounce by solving the Wheeler-DeWitt equation numerically with the Dirichlet or Neumann boundary condition in particular.

In the present paper, we will study the same system as in~\cite{ak} but under the Laplace-Beltrami operator-ordering and also with a more variety of boundary conditions.
In addition to the study of well-definedness of the quantum cosmology and the initial-singularity avoidance, we will also clarify whether the expectation value of the spatial volume of the universe obeys the classical time evolution in the late time.

The outline of the present paper is as follows.
In section~\ref{sec:ADM}, the ADM formalism with a Brown-Kucha\v{r} dust is reviewed.
In section~\ref{sec:QC}, we derive the Wheeler-DeWitt equation under the Laplace-Beltrami operator-ordering and determine the boundary condition for the wave function of the universe.
Section~\ref{sec:main} is devoted to studying the time evolution of a wave function.
Our results are summarized in section~\ref{sec:summary}.
A six-parameter family of exact solutions to the time-dependent Schr\"odinger equation for a free particle or a harmonic oscillator obtained in~\cite{lsv2013} is explained in appendix A, while exact time-dependent solutions on the half -line constructed with the Feynman kernel are presented in appendix B.
Our basic notation follows~\cite{wald}.
The convention for the Riemann curvature tensor is $[\nabla _\rho ,\nabla_\sigma]V^\mu ={R^\mu }_{\nu\rho\sigma}V^\nu$ and $R_{\mu \nu }={R^\rho }_{\mu \rho \nu }$.
The Minkowski metric is taken as diag$(-,+,+,+)$, and Greek indices run over all spacetime indices.
We adopt the units such that $c=1$.

\section{ADM formalism with a Brown-Kucha\v{r} dust}
\label{sec:ADM}
We consider general relativity in the presence of a cosmological constant $\Lambda$ in four dimensions, whose action is given by 
\begin{align}
S=\frac{1}{2\kappa^2}\int \D ^4x\sqrt{-g}(R-2\Lambda)+S_{\rm m}+S_{\partial{\cal M}},\label{action}
\end{align}
where $\kappa:=\sqrt{8\pi G}$ and $G$ is the Newton constant.
$S_{\rm m}$ is the action for matter fields and $S_{\partial{\cal M}}$ is the York-Gibbons-Hawking boundary term.
The resulting Einstein equations are
\begin{align} 
R_{\mu\nu}-\frac12 g_{\mu\nu}R+\Lambda g_{\mu\nu}=\kappa^2 T_{\mu\nu}, \label{beqL}
\end{align} 
where the energy-momentum tensor $T_{\mu\nu}$ is given from $S_{\rm m}$.

\subsection{Vacuum sector}

The most general four-dimensional metric may be written in the ADM form as
\begin{align}
\D s^2=&g_{\mu\nu}(x)\D x^\mu \D x^\nu \nonumber \\
=&-N^2\D t^2+h_{ij}(N^i\D t+\D x^i)(N^j\D t+\D x^j), \label{ADM}
\end{align}
where $N$, $N^i$, and $h_{ij}$ are functions of $t(\equiv x^0)$ and $x^i$($i=1,2,3$)~\cite{adm}.
Let $\Sigma$ a three-dimensional spacelike hypersurface with constant $t$ and then $h_{ij}$ is the induced metric on $\Sigma$.
In the present paper, we assume that $\Sigma$ is compact for simplicity.

In terms of the above ADM metric, the gravitational action is written as
\begin{align}
S_{\rm G}:=&\frac{1}{2\kappa^2}\int \D ^4x \sqrt{-g}({R}-2\Lambda) \nonumber \\
=&\int \D t \D ^3x\left(p^{ij}\partial_0 h_{ij}-NH-N_iH^i\right)+\mbox{(total derivative)}.
\end{align}
Here $p^{ij}$ is the momentum conjugate of $h_{ij}$ and the super-momentum $H^i$ and the super-Hamiltonian $H$ are respectively given by  
\begin{align}
H^i=&-2D_jp^{ij}, \label{m-const2} \\
H=&\frac{2\kappa^2}{\sqrt{h}}\biggl(p_{ij}p^{ij}-\frac{1}{2}p^2\biggl)-\frac{1}{2\kappa^2}\sqrt{h}({\cal R}-2\Lambda), \label{h-const2} 
\end{align}
where $h:=\mathrm{det}(h_{ij})$ and $p:=h_{ij}p^{ij}$.
$D_i$ and ${\cal R}$ are the covariant derivative and Ricci scalar on $\Sigma$, respectively.
The Lapse function $N$ and the shift vector $N_i$ act as the Lagrange multipliers and, in vacuum, the Euler-Lagrange equations for $N$ and $N_i$ give constraint equations $H=0$ and $H^i=0$, respectively.

\subsection{Matter sector}

In the present paper, we consider the Brown-Kucha\v{r} dust as a matter field~\cite{bk1995}.
It is a set of non-canonical scalar fields $\rho$, $T$, $Z^a$, and $W_a~(a=1,2,3)$ which are equivalent to a single timelike dust fluid, as explained below.

The action for the Brown-Kucha\v{r} dust is 
\begin{align}
S_{\rm m}=-\frac12\int \D ^4x\sqrt{-g}\rho(g^{\mu\nu}U_\mu U_\nu+1),
\end{align}
where $\rho$ represents the rest mass density and the one-form $U_\mu$ is defined by 
\begin{align}
U_\mu :=-(\nabla_\mu T)+W_a (\nabla_\mu Z^a).
\end{align}
The Euler-Lagrange equations corresponding to $T$, $Z^a$, and $W_a$ are
\begin{align}
\nabla_\mu (\rho U^\mu )=0,\qquad \rho  U^\mu (\nabla_\mu Z^a)=0,\qquad \nabla_\mu(\rho W_a U^\mu)=0, \label{eqU3}
\end{align}
respectively, while the Euler-Lagrange equation corresponding to $\rho$ is $g^{\mu\nu}U_\mu U_\nu=-1$.
The variation $\delta g^{\mu\nu}$ gives the following energy-momentum tensor;
\begin{align}
T_{\mu\nu}=\rho U_\mu U_\nu. \label{em-bk}
\end{align}
where we have used $g^{\mu\nu}U_\mu U_\nu=-1$.
From the Bianchi identity $\nabla_\nu (G^{\mu\nu}+\Lambda g^{\mu\nu})=0$, the energy-momentum conservation equations $\nabla_\nu T^{\mu\nu}=\nabla_\nu(\rho U^\mu U^\nu)=0$ hold.

In terms of the ADM metric, the dynamical part of the matter action is written as
\begin{align}
S_{\rm m}=\int \D t\D ^3x\biggl(P\partial_0 T+P_a (\partial_0 Z^a)-N^iH_i^{\rm D}-NH^{\rm D}\biggl), \label{BK-dust-ADM}
\end{align}
where $P$ and $P_a$ are the momentum conjugates of $T$ and $Z^a$, respectively~\cite{bk1995}.
The super-momentum $H_i^{\rm D}$ and super-Hamiltonian $H^{\rm D}$ for the Brown-Kucha\v{r} dust are respectively given by 
\begin{align}
H_i^{\rm D}=&P(\nabla_i T)+P_a(\nabla_i Z^a), \label{dustHi}\\
H^{\rm D}=&\sqrt{P^2+h^{ij}H_i^{\rm D}H_j^{\rm D}}.\label{dustH2}
\end{align}
The variables $W_a$ and $\rho$ are related to other ones as 
\begin{align}
W_a=&-P^{-1}P_a,\\
\frac{1}{\rho}=&\sqrt{h}|P|^{-1}\sqrt{P^{-2}h^{ij}H_i^{\rm D}H_j^{\rm D}+1}
\end{align}
and do not appear in the ADM form of the action (\ref{BK-dust-ADM}).

\subsection{FLRW minisuperspace}
 
In the present paper, we will study the FLRW quantum cosmology.
The line element of the FLRW spacetime is given by 
\begin{align}
\D s^2=-N(t)^2\D t^2+a(t)^2\gamma_{ij}\D x^i\D x^j,
\end{align}
where 
\begin{align}
\gamma_{ij}\D x^i\D x^j=\frac{\D r^2}{1-kr^2}+r^2(\D \theta^2+\sin^2\theta \D \varphi^2)
\end{align}
and $k=1,0,-1$ represents the spatial curvature of the universe.
We have assumed the compactness of the spatial section of the spacetime and then the spatial volume of the universe is given by $V(t)=V_0a(t)^3$, where $V_0:=\int \sqrt{\det (\gamma_{ij})}\D ^3x$.

We compute
\begin{align}
S_{\rm G}=&\frac{1}{2\kappa^2}\int \D ^4x \sqrt{-g}({R}-2\Lambda) \nonumber \\
=&\frac{V_0}{2\kappa^2}\int \D t \biggl\{\partial_t(6a^2N^{-1}{\dot a})+6Na^3\biggl(\frac{k}{a^2}-N^{-2}\frac{{\dot a}^2}{a^2}\biggl)-2\Lambda Na^3\biggl\},
\end{align}
where a dot denotes the derivative with respect to $t$.
Hence, the effective vacuum action to give the Einstein equations for the FLRW spacetime is  
\begin{align}
S_{\rm G}=\int L_{\rm G} \D t,
\end{align}
where 
\begin{align}
L_{\rm G}=\frac{3V_0}{\kappa^2}Na^{3}\biggl[-\frac{\Lambda}{3}+\frac{1}{N^2}\biggl(-\frac{{\dot a}^2}{a^2}+\frac{kN^2}{a^2}\biggl)\biggl].
\end{align}
The momentum conjugate of the scale factor $a$ and the super-Hamiltonian $H^{\rm G}$ are respectively given by
\begin{align}
p=&\frac{\delta L_{\rm G}}{\delta {\dot a}}= -\frac{6V_0}{\kappa^2}N^{-1}a{\dot a},\\
H^{\rm G}=&-\frac{\delta L_{\rm G}}{\delta N}=\frac{1}{2\kappa^2}\biggl(-\frac{\kappa^4}{6V_0}a^{-1}p^2+2V_0\Lambda a^{3}-6V_0ka\biggl). \label{HG}
\end{align}
Using them, we write the gravitational action in the ADM form:
\begin{align}
S_{\rm G}=\int (p{\dot a}-NH^{\rm G}) \D t.
\end{align}

For the matter sector, we consistently assume $Z^a \equiv 0$, $W_a \equiv 0$, $\rho=\rho(t)$, and $T=T(t)$.
Then we have $H_i^{\rm D}=0$ and $H^{\rm D}=P$, where we have taken the plus sign.
Finally, the total action in the ADM form is given by  
\begin{align}
S=S_{\rm G}+S_{\rm m}=\int\biggl(p{\dot a}+P{\dot T}-N(H^{\rm G}+H^{\rm D})\biggl) \D t. \label{ADM-FRW}
\end{align}
Now the Hamiltonian constraint $H^{\rm G}+H^{\rm D}=0$ is written as
\begin{align}
\frac{1}{2\kappa^2}\biggl(-\frac{\kappa^4}{6V_0}a^{-1}p^2+2V_0\Lambda a^{3}-6V_0ka\biggl)+P=0.\label{h-basic}
\end{align}
It is noted that Amemiya and Koike considered a radiation fluid in addition in their analysis~\cite{ak}.
In the present paper, we don't consider such an additional matter field for simplicity in order to pursue exact results as much as possible.

The Euler-Lagrange equation for $T(t)$ is ${\ddot T}=0$, which is solved to give $T\propto t-t_0$, where $t_0$ is a constant.
For this reason, we may use the scalar field $T$ as a clock in the present system.
In the comoving coordinates $U_\mu\D x^\mu =-\D t$ corresponding to $T=t-t_0$, the classical solution in this system for $k=0$ with $\Lambda=0$ is $a(t)\propto t^{2/3}$.
In the presence of positive $\Lambda$, the late-time behavior of the scale factor is $\lim_{t\to\infty}a(t)\propto e^{\sqrt{\Lambda/3}t}$.
We will see that $T$ appears as time in the corresponding quantum system.

\section{Quantum cosmology}
\label{sec:QC}
\subsection{Laplace-Beltrami operator-ordering}
We quantize the system (\ref{ADM-FRW}) by replacing the momentum conjugates by operators as $p\to {\hat p}=-i\hbar \delta/\delta a$ and $P\to {\hat P}=-i\hbar (\delta/\delta T)$ and then the Hamiltonian constraint (\ref{h-basic}) gives the following Wheeler-DeWitt equation:
\begin{align}
{\hat H}^{\rm G}\Psi=i\hbar \frac{\delta \Psi}{\delta T}, \label{WdW}
\end{align}
where $\Psi=\Psi[a,T]$ is the wave function(al) of the universe.
This is the form of the Schr\"odinger equation where the scalar field $T$ acts as a time variable.
Since ${\hat H}^{\rm G}$ is an operator obtained from Eq.~(\ref{HG}), there is an ambiguity of the operator-ordering.
In the present paper, we adopt the Laplace-Beltrami operator-ordering~\cite{zanelli1986}, which is natural in the following sense.

By way of explanation, let us consider spatially homogeneous and anisotropic cosmological models to be quantized. 
According to the Bianchi classification, such spacetimes have three dynamical degrees of freedom at most, which we denote by $X^I(t)~(I=1,2,3)$.
In this Bianchi minisuperspace, classical systems are equivalent to the dynamics of a point particle in a curved space.
Then in general, the super-Hamiltonian (\ref{HG}) can be written in the following form:
\begin{align}
H^{\rm G}=&\frac{1}{2m}{\cal G}^{IJ}[X]p_Ip_J+V[X],
\end{align}
where $m$ is the effective mass, $p_I$ is the momentum conjugate of $X^I$, and $V$ is the effective potential. 
From this expression, we can read off ${\cal G}^{IJ}$, the contravariant components of the supermetric ${\cal G}_{IJ}$ in the DeWitt superspace.
By analogy with the canonical quantization of a point particle, the Laplace-Beltrami operator-ordering then requires ${\hat H}^{\rm G}$ to be
\begin{align}
{\hat H}^{\rm G}=&-\frac{\hbar^2}{2m}\Delta_{\cal G}+V[X].
\end{align}
Here $\Delta_{\cal G}$ is the Laplacian in the superspace:
\begin{align}
\Delta_{\cal G}:=\frac{1}{\sqrt{-{\cal G}}}\frac{\delta}{\delta X^I}\biggl(\sqrt{-{\cal G}}{\cal G}^{IJ}\frac{\delta}{\delta X^J}\biggl),
\end{align}
where ${\cal G}:=\det ({\cal G}_{IJ})$.
A natural inner product under this operator-ordering is  
\begin{align}
\langle \Phi|\Psi\rangle:=\int \Phi^\ast \Psi \sqrt{-{\cal G}}\D^3X.
\end{align}

\subsection{Wheeler-DeWitt equation}

The Laplace-Beltrami operator-ordering leads
\begin{align}
{\hat H}^{\rm G}=\frac{1}{2\kappa^2}\biggl(-\frac{\kappa^4}{6V_0}a^{-1/2}{\hat p} a^{-1/2}{\hat p}+2V_0\Lambda a^{3}-6V_0ka\biggl)
\end{align}
and finally the Wheeler-DeWitt equation (\ref{WdW}) becomes
\begin{align}
\biggl(\frac{3\kappa^2\hbar^2}{16V_0}\frac{\delta^2}{\delta x^2}+\frac{V_0\Lambda}{\kappa^2} x^2-\frac{3V_0k}{\kappa^2}x^{2/3}\biggl)\Psi=i\hbar \frac{\delta\Psi}{\delta T}, \label{WdW2}
\end{align}
where $x:=a^{3/2}$.
The domain of $a$ (and also $x$) is $[0,\infty)$. 
In terms of $x$, the inner product is simply 
\begin{align}
\langle \Phi|\Psi\rangle=\int^\infty_0\Phi^*\Psi\D{x}. \label{naiseki}
\end{align}  

The equivalent Schr\"odinger equation to the Wheeler-DeWitt equation (\ref{WdW2}) is 
\begin{align}
\biggl(-\frac{\hbar^2}{2m}\frac{\partial^2}{\partial x^2}+V(x)\biggl)\Psi=i\hbar \frac{\partial \Psi}{\partial T}, \label{Schro-eq}
\end{align}  
where the effective potential $V(x)$ and mass $m$ are
\begin{align}
V(x)=&\frac{V_0\Lambda}{\kappa^2} x^2-\frac{3V_0k}{\kappa^2}x^{2/3},\label{p-potential}\\
m=&-\frac{8V_0}{3\kappa^2}. \label{p-mass}
\end{align}  
Now our problem has reduced to quantum mechanics on the half-line and we will study the one-dimensional Schr\"odinger equation (\ref{Schro-eq}) with the inner product (\ref{naiseki}).
Hereafter we will consider only the spatially flat case $k=0$ for simplicity, in which exact solutions are available.

\subsection{Boundary condition for the wave function}
A quantum system is well-defined if the Hamiltonian operator 
\begin{align}
{\hat H}^{\rm G}=-\frac{\hbar^2}{2m}\frac{\partial^2}{\partial x^2}+V(x)
\end{align}
is self-adjoint.
Then, by the Stone's theorem on one-parameter unitary groups, the time evolution of a wave function is uniquely determined by 
\begin{align}
\Psi(x,T)=e^{-i{\hat H}^{\rm G}T/\hbar}\Psi(x,0). \label{evolution}
\end{align}
Let us define the Hamiltonian operator ${\hat H}^{\rm G}$ on a dense domain ${\cal D}({\hat H}^{\rm G})=C_0^\infty((0,\infty))$, the space of smooth functions compactly supported in $(0,\infty)$, within the square-integrable Hilbert space ${\cal L}^2((0,\infty))$.
Actually, ${\hat H}^{\rm G}$ is not self-adjoint because the domain of its adjoint operator ${\hat H}^{{\rm G}\dagger}$ is strictly larger than ${\cal D}({\hat H}^{\rm G})$.
However, we may consider self-adjoint extensions of ${\hat H}^{\rm G}$ and then the time evolution is given by Eq.~(\ref{evolution}) in their domains.
(See~\cite{Reed-Simon,robin bcs} for a textbook and reviews.)
We are going to show that our Hamiltonian operator ${\hat H}^{\rm G}$ with $k=0$ has self-adjoint extensions characterized by one real parameter, namely it admits an infinite number of self-adjoint extensions.

For this purpose, we consider the Hilbert space eigenproblem ${\hat H}^{{\rm G}\dagger}\Psi=\pm i\Psi$ for $\Psi \in {\cal L}^2((0,\infty))$. 
Based on the results in Section X.1 in~\cite{Reed-Simon}, this problem reduces to solving the ordinary differential equations ${\hat H}^{{\rm G}\dagger}\Psi=\pm i\Psi$ for smooth square-integrable functions $\Psi$.
The deficiency indices $n_{\pm}$ are the dimensions of the space of general solutions to these differential equations without imposing boundary conditions at $x=0$. 
Namely, $n_+$ and $n_-$ denote the numbers of parameters contained in the general solutions to the differential equations with $+i$ and $-i$, respectively.
If $n_{\pm}=0$, then ${\hat H}^{\rm G}$ is essentially self-adjoint and no further boundary conditions are required.
If $n_+ \ne n_-$, then ${\hat H}^{\rm G}$ has no self-adjoint extensions and the quantum system is ill-defined.
If $n_+=n_- \ne 0$, then ${\hat H}^{\rm G}$ has self-adjoint extensions which require the imposition of
further boundary conditions with $n_+=n_-$ parameters.

In the case of $\Lambda=k=0$, the general solutions to the differential equations ${\hat H}^{\rm G\dagger}\Psi=\pm i\Psi$ are
\begin{align}
\Psi(x)=C_1\exp\biggl(\sqrt{\frac{16V_0}{3\kappa^2\hbar^2}}\frac{1\pm i}{\sqrt{2}}x\biggl)+C_2\exp\biggl(-\sqrt{\frac{16V_0}{3\kappa^2\hbar^2}}\frac{1\pm i}{\sqrt{2}}x\biggl).
\end{align}
Independent of the sign in the differential equation, the term with $C_1$ diverges for $x\to \infty$ and only the term with $C_2$ provides a solution in ${\cal L}^2((0,\infty))$.
This shows that the deficiency indices are $n_+=n_-=1$ and therefore our Hamiltonian operator ${\hat H}^{\rm G}$ with $\Lambda=k=0$ has an infinite number of self-adjoint extensions characterized by one parameter.

The result is the same also with a positive cosmological constant ($\Lambda>0$).
In this case, by the scaling transformations $z:=\beta{x}$ and ${\bar T}:=(3\kappa^2\hbar \beta^2/16V_0)T$ with $\beta:=(64V_0^2\Lambda/3\kappa^4\hbar^2)^{1/4}$, the Schr\"odinger equation (\ref{Schro-eq}) becomes ${\hat H}\Psi=i \partial \Psi/\partial {\bar T}$, where 
\begin{align}
{\hat H}:=\frac{\partial^2}{\partial z^2}+\frac14  z^2.
\end{align}  
Hence, we consider the following ordinary differential equations
\begin{align}
\frac{\partial^2\Psi}{\partial z^2}+\frac14  z^2\Psi=\pm i\Psi
\end{align}  
 for smooth square-integrable functions $\Psi(z)$ without imposing boundary conditions at $z=0$ (and hence $x=0$).
The general solutions to these differential equations are 
\begin{align}
\Psi(z)=z^{3/2}\biggl\{C_3\biggl(I_{3/4}(\pm iz^2/4)+I_{-1/4}(\pm iz^2/4)\biggl)+C_4\biggl(K_{3/4}(\pm iz^2/4)-K_{1/4}(\pm iz^2/4)\biggl)\biggl\},
\end{align}  
where $C_3$ and $C_4$ are constants and $I_\alpha(x)$ and $K_\alpha(x)$ are the modified Bessel functions of the first and second kinds, respectively. (See Section 12.14 in~\cite{NIST}.) 
Clearly, the term with the constant $C_3$ diverges for $z\to \infty$ and only the term with the constant $C_4$ provides a solution in ${\cal L}^2((0,\infty))$.

We have shown that our Hamiltonian operator ${\hat H}^{\rm G}$ with $k=0$ has self-adjoint extensions which require a boundary condition with one parameter.
This boundary condition is determined so as to satisfy the symmetric property ${\hat H}^{{\rm G}\dagger}={\hat H}^{\rm G}$. 
Using the Schr\"odinger equation (\ref{Schro-eq}) and integration by parts together with the fall-off condition at infinity, we obtain
\begin{align}
\langle \Phi|{\hat H}^{\rm G}\Psi\rangle=\frac{\hbar^2}{2m}\biggl(\Phi^*\frac{\partial\Psi}{\partial x}-\frac{\partial\Phi^*}{\partial x}\Psi\biggl)\biggl|_{x=0}+\langle {\hat H}^{\rm G}\Phi|\Psi\rangle.
\end{align}  
In order for the surface term to be vanishing, we impose boundary conditions on $\Phi$ and $\Psi$ such as
\begin{align}
&\Phi(0,T)+L\frac{\partial \Phi}{\partial  {x}}(0,T)=0,\\
&\Psi(0,T)+L\frac{\partial \Psi}{\partial  {x}}(0,T)=0, \label{b-condition}
\end{align}  
where $L$ is a real constant, and then ${\hat H}^{{\rm G}\dagger}={\hat H}^{\rm G}$ is realized.
These boundary conditions also emerge from the deficiency space machinery as discussed in~\cite{Reed-Simon}.

Since the boundary condition (\ref{b-condition}) contains one real parameter $L$, the Hamiltonian operator ${\hat H}^{\rm G}$ admits an infinite number of self-adjoint extensions and each value of $L$ gives a different quantum system.
This boundary condition ensures unitarity $\partial\langle \Psi|\Psi\rangle/\partial T=0$, shown as
\begin{align}
\frac{\partial}{\partial T}\langle \Psi|\Psi\rangle=\frac{i\hbar}{2m}\biggl(\frac{\partial\Psi^*}{\partial x}\Psi-\Psi^*\frac{\partial\Psi}{\partial x}\biggl)\biggl|_{x=0}=0.
\end{align}  
$L=0$ and $L=\infty$ correspond to the Dirichlet and Neumann boundary conditions at $x=0$, respectively, and other values of $L$ correspond to the Robin boundary condition.

\section{Unitary evolution of the quantum universe}
\label{sec:main}
In this section, we will solve the the Schr\"odinger equation~(\ref{Schro-eq}) (which is equivalent to the Wheeler-DeWitt equation (\ref{WdW2})) with $k=0$ and see dynamical properties of a wave function depending on the extension parameter $L$ and $\Lambda$.
In particular, we will check whether the expectation value of the spatial volume of the universe, which is proportional to $\langle a^3\rangle(=\langle x^2\rangle)$, obeys the classical time evolution $\langle a^3\rangle\propto T^{2}$ (for $\Lambda=0$) or $\langle a^3\rangle\propto \exp(\sqrt{3\Lambda}T)$ (for $\Lambda>0$) in the late time.

In the case of the full-line $x\in (-\infty,\infty)$, the following Ehrenfest's theorem holds;
\begin{align}
m\frac{\partial^2\langle x\rangle}{\partial T^2}=-\biggl\langle \frac{\partial V}{\partial x}\biggl\rangle,
\end{align}
and therefore $\langle x\rangle$ follows classical orbits.
The above equation is modified in the half-line case $x\in [0,\infty)$.
Using the Schr\"odinger equation~(\ref{Schro-eq}) and integration by parts, we obtain
\begin{align}
m\frac{\partial^2\langle x^q\rangle}{\partial T^2}
=&\frac{\hbar}{2i}\frac{\partial}{\partial T}\biggl[x^q\biggl(\frac{\partial\Psi^*}{\partial x} \Psi-\Psi^*\frac{\partial\Psi}{\partial x}\biggl)\biggl]_0^\infty  \nonumber \\
&+ \frac{q\hbar^2}{4m}\biggl[x^{q-1}\biggl(\frac{\partial^2\Psi^*}{\partial x^2}\Psi+\Psi^*\frac{\partial^2\Psi}{\partial x^2}-2\frac{\partial\Psi^*}{\partial x}\frac{\partial\Psi}{\partial x}\biggl)-(q-1)x^{q-2}\biggl(\frac{\partial\Psi^*}{\partial x}\Psi+\Psi^*\frac{\partial\Psi}{\partial x}\biggl) \biggl]_0^\infty \nonumber \\
&-\int_0^\infty qx^{q-1}\Psi^*\frac{\partial V}{\partial x}\Psi \D x \nonumber \\
&+\frac{q(q-1)\hbar^2}{4m}\int_0^\infty \biggl((q-2)x^{q-3}\frac{\partial(\Psi^*\Psi)}{\partial x}+4x^{q-2}\frac{\partial\Psi^*}{\partial x}\frac{\partial\Psi}{\partial x}\biggl) \D x \label{e-theorem1}
\end{align}
for constant $q$.
The surface term at $x=0$ in the first term vanishes for positive $q$ by the boundary condition (\ref{b-condition}).
Then, under the assumption that the surface terms at infinity vanish, Eq.~(\ref{e-theorem1}) reduces to
\begin{align}
m\frac{\partial^2\langle x^q\rangle}{\partial T^2}
=& -\frac{q\hbar^2}{4m}\biggl[x^{q-1}\biggl(\frac{\partial^2\Psi^*}{\partial x^2}\Psi+\Psi^*\frac{\partial^2\Psi}{\partial x^2}-2\frac{\partial\Psi^*}{\partial x}\frac{\partial\Psi}{\partial x}\biggl)-(q-1)x^{q-2}\biggl(\frac{\partial\Psi^*}{\partial x}\Psi+\Psi^*\frac{\partial\Psi}{\partial x}\biggl) \biggl]\biggl|_{x=0} \nonumber \\
&-\int_0^\infty qx^{q-1}\Psi^*\frac{\partial V}{\partial x}\Psi \D x \nonumber \\
&+\frac{q(q-1)\hbar^2}{4m}\int_0^\infty \biggl((q-2)x^{q-3}\frac{\partial(\Psi^*\Psi)}{\partial x}+4x^{q-2}\frac{\partial\Psi^*}{\partial x}\frac{\partial\Psi}{\partial x}\biggl) \D x
\end{align}
for positive $q$.
For $q=1$, this becomes
\begin{align}
m\frac{\partial^2\langle x\rangle}{\partial T^2}=&-\frac{\hbar^2}{4m} \biggl(\frac{\partial^2\Psi^*}{\partial x^2}\Psi+\Psi^*\frac{\partial^2\Psi}{\partial x^2}-2\frac{\partial\Psi^*}{\partial x}\frac{\partial\Psi}{\partial x}\biggl)\biggl|_{x=0} -\biggl\langle \frac{\partial V}{\partial x}\biggl\rangle
\end{align}
and therefore in general, the expectation value of $x$ does not follow classical orbits.

Our interest is the evolution of $\langle a^3\rangle(=\langle x^2\rangle)$ and this quantity satisfies the following equation:
\begin{align}
m\frac{\partial^2\langle x^2\rangle}{\partial T^2}
=& -\frac{\hbar^2}{2m}\biggl[x\biggl(\frac{\partial^2\Psi^*}{\partial x^2}\Psi+\Psi^*\frac{\partial^2\Psi}{\partial x^2}-2\frac{\partial\Psi^*}{\partial x}\frac{\partial\Psi}{\partial x}\biggl)-\biggl(\frac{\partial\Psi^*}{\partial x}\Psi+\Psi^*\frac{\partial\Psi}{\partial x}\biggl) \biggl]\biggl|_{x=0} \nonumber \\
&+\int_0^\infty \biggl(\frac{2\hbar^2}{m}\frac{\partial\Psi^*}{\partial x}\frac{\partial\Psi}{\partial x}-2x\Psi^*\frac{\partial V}{\partial x}\Psi \biggl) \D x.
\end{align}

\subsection{Analytical results}

\subsubsection{Exact wave function I}

As seen in Eq.~(\ref{Schro-eq}), the Wheeler-DeWitt equation~(\ref{WdW2}) with $\Lambda>0$ and $k=0$ is equivalent to the Schr\"odinger equation for a harmonic oscillator with negative mass.
By the transformation $x=\sqrt{3\kappa^2\hbar/(8V_0)}{\bar x}$, Eq.~(\ref{Schro-eq}) with $k=0$ becomes
\begin{align}
\frac12\frac{\partial^2 \Psi}{\partial {\bar x}^2}+\frac12{\bar k}^2{\bar x}^2\Psi=i\frac{\partial \Psi}{\partial {T}}, \label{schr1}
\end{align}  
where ${\bar k}^2:=3\Lambda/4$.
In the case of the Dirichlet or Neumann boundary condition at the origin, exact time-dependent solutions to Eq.~(\ref{schr1}) are available.

First, we use the six-parameter family of exact solutions obtained in~\cite{lsv2013}.
(See Appendix~\ref{app:six}.)
The solution to Eq.~(\ref{schr1}) is obtained by $x\to {\bar x}$ and $t\to -T$ from Eqs.~(\ref{six-solution}) and (\ref{six-anti1})--(\ref{six-anti2}) as
\begin{align}
\Psi({\bar x},T)=&\Psi_n({\bar x},T) \nonumber \\
:=&\frac{e^{i\left(\alpha(T){\bar x}^2+\delta(T){\bar x}+\kappa(T)\right)+i(2n+1)\gamma(T)}}{\sqrt{2^nn!\mu(T)\sqrt{\pi}}}e^{-(\beta(T){\bar x}+\varepsilon(T))^2/2}H_n\left(\beta(T){\bar x}+\varepsilon(T)\right), \label{exact-I}
\end{align}
where $H_n(x)~(n=0,1,2,\cdots)$ is the Hermite polynomials and  
\begin{align}
\mu(T)=&\mu_0\sqrt{{\bar\beta}_0^4\sinh^2{\bar k}T/{\bar k}^2+(2{\bar\alpha}_0\sinh{\bar k}T-\cosh{\bar k}T)^2},\\
\alpha(T)=&\frac{{\bar k}{\bar\alpha}_0\cosh 2{\bar k}T-(\sinh 2{\bar k}T/{\bar k})({\bar\beta}_0^4-4{\bar k}^2\alpha_0^2+{\bar k}^2)/4}{{\bar\beta}_0^4\sinh^2{\bar k}T/{\bar k}^2+(2{\bar\alpha}_0\sinh{\bar k}T-\cosh{\bar k}T)^2},\\
\beta(T)=&\frac{{\bar\beta}_0}{\sqrt{{\bar\beta}_0^4\sinh^2{\bar k}T/{\bar k}^2+(2{\bar\alpha}_0\sinh{\bar k}T-\cosh{\bar k}T)^2}},\\
\gamma(T)=&\gamma_0-\frac12\arctan\biggl(\frac{{\bar\beta}_0^2\sinh {\bar k}T/{\bar k}}{2{\bar\alpha}_0\sinh{\bar k}T-\cosh{\bar k}T}\biggl),\\
\delta(T)=&-\frac{{\bar\delta}_0(2{\bar\alpha}_0\sinh{\bar k}T-\cosh{\bar k}T)+\varepsilon_0{\bar\beta}_0^3\sinh {\bar k}T/{\bar k}}{{\bar\beta}_0^4\sinh^2{\bar k}T/{\bar k}^2+(2{\bar\alpha}_0\sinh{\bar k}T-\cosh{\bar k}T)^2},\\
\varepsilon(T)=&-\frac{\varepsilon_0(2{\bar\alpha}_0\sinh{\bar k}T-\cosh{\bar k}T)-{\bar\beta}_0{\bar\delta}_0\sinh {\bar k}T/{\bar k}}{\sqrt{{\bar\beta}_0^4\sinh^2{\bar k}T/{\bar k}^2+(2{\bar\alpha}_0\sinh{\bar k}T-\cosh{\bar k}T)^2}},\\
\kappa(T)=&\kappa_0+\frac{\sinh^2 {\bar k}T}{{\bar k}^2}\frac{\varepsilon_0{\bar\beta}_0^2({\bar k}{\bar \alpha}_0\varepsilon_0-{\bar\beta}_0{\bar\delta}_0)-{\bar k}{\bar \alpha}_0{\bar\delta}_0^2}{{\bar\beta}_0^4\sinh^2{\bar k}T/{\bar k}^2+(2{\bar\alpha}_0\sinh{\bar k}T-\cosh{\bar k}T)^2} \nonumber \\
&-\frac14\frac{\sinh 2{\bar k}T}{{\bar k}}\frac{\varepsilon_0^2{\bar\beta}_0^2-{\bar\delta}_0^2}{{\bar\beta}_0^4\sinh^2{\bar k}T/{\bar k}^2+(2{\bar\alpha}_0\sinh{\bar k}T-\cosh{\bar k}T)^2}.
\end{align}
One of the seven parameters $\mu_0,{\bar \alpha}_0,{\bar \beta}_0,\gamma_0,{\bar\delta}_0,\kappa_0,\varepsilon_0$ may be used for normalization, so that the number of independent parameters is six.
In the free-particle limit (${\bar k}\to 0$), we have
\begin{align}
\mu(T)=&\mu_0\sqrt{{\bar\beta}_0^4T^2+1},\qquad \alpha(T)=-\frac{{\bar\beta}_0^4T}{2({\bar\beta}_0^4T^2+1)},\\
\beta(T)=&\frac{{\bar\beta}_0}{\sqrt{{\bar\beta}_0^4T^2+1}},\qquad \gamma(T)=\gamma_0+\frac12\arctan({\bar\beta}_0^2T),\\
\delta(T)=&\frac{{\bar\delta}_0-\varepsilon_0{\bar\beta}_0^3T}{{\bar\beta}_0^4T^2+1},\qquad \varepsilon(T)=\frac{\varepsilon_0+{\bar\beta}_0{\bar\delta}_0T}{\sqrt{{\bar\beta}_0^4T^2+1}},\\
\kappa(T)=&\kappa_0-\frac{\varepsilon_0{\bar\beta}_0^3{\bar\delta}_0T^2}{{\bar\beta}_0^4T^2+1} -\frac{(\varepsilon_0^2{\bar\beta}_0^2-{\bar\delta}_0^2)T}{2({\bar\beta}_0^4T^2+1)},
\end{align}
where the parameter ${\bar \alpha}_0$ has disappeared.

\begin{figure}[htbp]
\begin{center}
\includegraphics[width=1.0\linewidth]{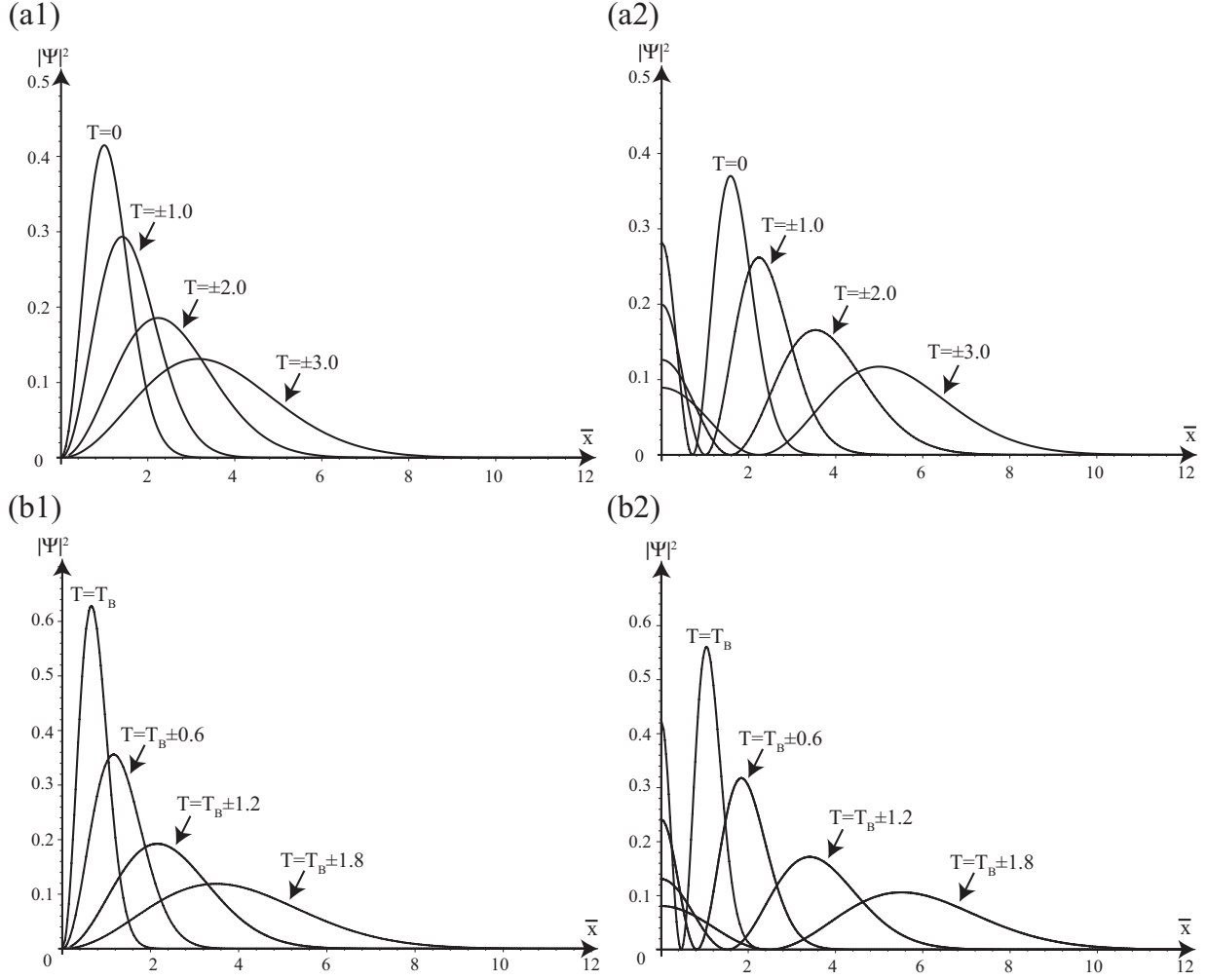}
\caption{\label{wave-exact} Time evolution of $|\Psi|^2$ for the exact wave function (\ref{exact-I}) with (a1) $\Lambda=0$ and $n=1$, (a2) $\Lambda=0$ and $n=2$, (b1) $\Lambda=0.5$ and $n=1$, (b2) $\Lambda=0.5$ and $n=2$, where we set $\mu_0=1$, ${\bar \alpha}_0=1$, ${\bar \beta}_0=1$, $\gamma_0=1$, ${\bar\delta}_0=0$, $\kappa_0=1$, and $\varepsilon_0=0$. The big bounce occurs at $T=T_{\rm B}$ and $|\Psi|^2$ is time-symmetric with respect to $T_{\rm B}$, where $T_{\rm B}$ is defined by Eq.~(\ref{TB}). 
}
\end{center}
\end{figure}

This family of solutions with $\varepsilon_0={\bar\delta}_0=0$ (hence $\varepsilon(T)=0$) are also solutions in ${\cal L}^2([0,\infty))$ with the Dirichlet or Neumann boundary condition at the origin. 
The Dirichlet boundary condition is satisfied for odd $n$, while the Neumann boundary condition is satisfied for even $n$.
The solutions are regular everywhere for $T\in (-\infty,\infty)$ and hence the classical initial singularity is avoided in the corresponding quantum system.
Figure~\ref{wave-exact} shows the evolution of $|\Psi|^2$ for $n=1$ and $n=2$.

Let us discuss the late-time evolution of the expectation value of the spatial volume of the universe, which is proportional to $\langle a^3\rangle(=\langle x^2\rangle)$.
For $\varepsilon(T)=0$ ($\varepsilon_0={\bar\delta}_0=0$) and an integer $q$, we obtain
\begin{align}
\int_{0}^\infty {\bar x}^q|\Psi_n({\bar x},T)|^2\D {\bar x}=&\frac{1}{2^nn!\mu(T)\sqrt{\pi}}\int_{0}^\infty {\bar x}^q e^{-\beta(T)^2{\bar x}^2}H_n(\beta(T){\bar x})^2 \D {\bar x} \nonumber \\
=&\frac{1}{2^nn!\mu(T)\beta(T)^{1+q}\sqrt{\pi}}\int_{0}^{\pm\infty} y^q e^{-y^2}H_n(y)^2 \D y,
\end{align}
where the sign in $\pm\infty$ corresponds to the sign of ${\bar \beta}_0$.
Using this, we compute 
\begin{align}
\langle x^q\rangle \propto  \frac{\int_{0}^\infty {\bar x}^q|\Psi_n({\bar x},T)|^2\D {\bar x}}{\int_{0}^\infty |\Psi_n({\bar x},T)|^2\D {\bar x}}=\frac{1}{2^{n-1}n!\beta(T)^q\sqrt{\pi}}\int_{0}^{\pm\infty} y^q e^{-y^2}H_n(y)^2 \D y, \label{x^q}
\end{align}
where we used 
\begin{align}
\int_{0}^{\infty} e^{-y^2}H_n(y)^2\D y=\sqrt{\pi}2^{n-1}n!.
\end{align}

\begin{figure}[htbp]
\begin{center}
\includegraphics[width=0.6\linewidth]{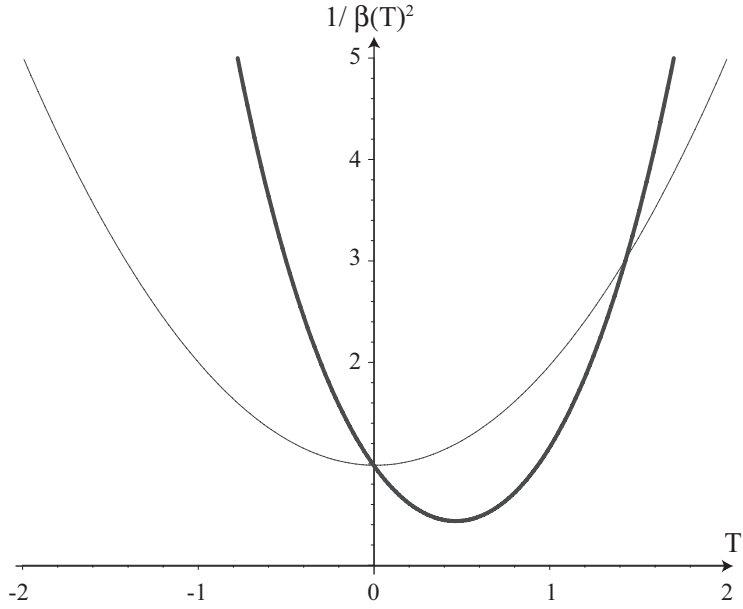}
\caption{\label{bounce-exact} The function $\beta(T)^{-2}(\propto \langle a^3\rangle)$ for $\Lambda=0$ (thin curve) and $\Lambda=0.5$ (thick curve), where we set ${\bar \alpha}_0=1$ and ${\bar \beta}_0=1$. 
The big bounce occurs at $T=T_{\rm B}$, defined by Eq.~(\ref{TB}), where $T_{\rm B}=0$ for $\Lambda=0$ and $T_{\rm B}\simeq 0.47253$ for $\Lambda=0.5$.
}
\end{center}
\end{figure}

Equation~(\ref{x^q}) shows that the time-dependence of $\langle a^3\rangle$ is given by 
\begin{align}
\langle a^3\rangle \propto \frac{1}{\beta(T)^2}=& \left\{
\begin{array}{ll}
\displaystyle{{\bar\beta}_0^{-2}({\bar\beta}_0^4T^2+1)} &\mbox{for} \quad \Lambda=0,\\
\displaystyle{{\bar\beta}_0^{-2}\biggl({\bar\beta}_0^4{\bar k}^{-2}\sinh^2{\bar k}T+(2{\bar\alpha}_0\sinh{\bar k}T-\cosh{\bar k}T)^2\biggl)} & \mbox{for}\quad \Lambda>0.
\end{array} \right. 
\end{align}
In both cases, $\langle a^3\rangle$ is positive definite and it admits only one local minimum at $T=T_{\rm B}$, where
\begin{align}
\label{TB}
T_{\rm B}=& \left\{
\begin{array}{ll}
\displaystyle{0} &\mbox{for} \quad \Lambda=0,\\
\displaystyle{\frac{1}{4{\bar k}}\ln\biggl(\frac{{\bar\beta}_0^4+{\bar k}^2(1+2{\bar\alpha}_0)^2}{{\bar\beta}_0^4+{\bar k}^2(1-2{\bar\alpha}_0)^2}\biggl)} & \mbox{for}\quad \Lambda>0,
\end{array} \right. 
\end{align}
corresponding to the transition time from the contracting phase ($\D\langle a^3\rangle/\D T<0$) to the expanding phase ($\D\langle a^3\rangle/\D T>0$).
Thus, the classical initial singularity is avoided and replaced by the big bounce at $T=T_{\rm B}$ in the quantum system.
Figure~\ref{bounce-exact} shows the function $\beta(T)^{-2}(\propto \langle a^3\rangle)$.

In addition, the late-time evolution of $\langle a^3\rangle$ is
\begin{align}
\lim_{T\to \infty}\langle a^3 \rangle\propto & \left\{
\begin{array}{ll}
\displaystyle{T^2} &\mbox{for} \quad \Lambda=0,\\
\displaystyle{e^{2{\bar k}T}} & \mbox{for}\quad \Lambda>0.
\end{array} \right. 
\end{align}
Since we have ${\bar k}=\sqrt{3\Lambda/4}$ in our system, this proves the convergence to the classical evolution in both cases.
On the other hand, the variance of $a^3$ diverges as 
\begin{align}
\lim_{T\to \infty}{\rm Var}(a^3)=\lim_{T\to \infty}\biggl(\langle a^6 \rangle-\langle a^3 \rangle^2\biggl) \propto \lim_{T\to \infty}\beta(T)^{-4}\to \infty.
\end{align}

\subsubsection{Exact wave function II}
To the Schr\"odinger equation (\ref{Schro-eq}) with $\Lambda>0$ and $k=0$, there is also another class of exact solutions with Dirichlet or Neumann boundary condition, constructed by using the Feynman kernel.
(See Appendix~\ref{app:kernel} for derivation.)

In the case of ${\cal L}^2((-\infty,\infty))$, the typical initial profile is the following Gaussian wave packet:
\begin{align}
\Psi(x,0)=C\exp\biggl(-\frac{(x-x_0)^2}{4\sigma^2}+i\frac{p_0x}{\hbar}\biggl), \label{boundary-i}
\end{align}  
where $p_0$, $x_0$, and $\sigma$ are constants and the normalization constant $C$ is $1/(2\pi \sigma^2)^{1/4}$.
This wave function represents a moving wave packet with its peak at $x=x_0$ and its momentum $p$, as shown by $\langle x\rangle(0)=x_0$ and $\langle p\rangle(0)=p_0$.

In the half-line case ${\cal L}^2([0,\infty))$, the physical meanings of the constants $x_0$ and $p_0$ are less clear and the profile (\ref{boundary-i}) does not satisfy neither the Dirichlet nor Neumann boundary condition at $x=0$.
From this observation, we modify the initial profile as
\begin{align}
\Psi(x,0)=x\exp\biggl(-\frac{(x-x_0)^2}{4\sigma^2}+i\frac{p_0x}{\hbar}\biggl) \label{initial1}
\end{align}   
for $L=0$ and 
\begin{align}
\Psi(x,0)=\biggl[x-\biggl\{\frac{1}{L}+\biggl(\frac{x_0}{2\sigma^2}+i\frac{p_0}{\hbar}\biggl)\biggl\}^{-1}\biggl]\exp\biggl(-\frac{(x-x_0)^2}{4\sigma^2}+i\frac{p_0x}{\hbar}\biggl)  \label{initial2}
\end{align}  
for $L\ne 0$, which satisfy the boundary condition (\ref{b-condition}).

The solution with the Dirichlet boundary condition and the initial profile (\ref{initial1}) is obtained from Eq.~(\ref{harmonic-d}) by the reparametrization $\omega\to i{\bar \omega}$ and identifying $m=-8V_0/3\kappa^2$ and ${\bar \omega}^2=3\Lambda/4$:
\begin{align}
\Psi(x,T)
=&\frac{1}{2\eta}\sqrt{\frac{m{\bar\omega}}{2\pi i\hbar \sinh{\bar\omega} T}}\exp\biggl(-\frac{x_0^2}{4\sigma^2}+\frac{im{\bar\omega} x^2\cosh{\bar\omega} T}{2\hbar\sinh{\bar\omega} T}\biggl)  \nonumber \\
&\times\biggl\{\frac{1}{2}\sqrt{\frac{\pi}{\eta }}\biggl(\zeta e^{\zeta^2/4\eta }-{\bar \zeta}e^{{\bar \zeta}^2/4\eta }\biggl)+i\zeta F\biggl(-\frac{i\zeta}{2\sqrt{\eta }}\biggl)-i{\bar \zeta} F\biggl(-\frac{i{\bar \zeta}}{2\sqrt{\eta }}\biggl)\biggl\}, \label{sol-d-harmonic}
\end{align} 
where $F(z)$ is the Dawson function defined by 
\begin{align}
F(z):=e^{-z^2}\int_0^ze^{w^2}\D w
\end{align} 
and $\eta=\eta(T)$, $\zeta=\zeta(x,T)$, and ${\bar \zeta}={\bar \zeta}(x,T)$ are given by 
\begin{align}
\begin{aligned}
\eta(T) =& \frac{1}{4\sigma^2}-\frac{im{\bar\omega} \cosh{\bar\omega} T}{2\hbar\sinh{\bar\omega} T},\\
\zeta(x,T) =&\frac{x_0}{2\sigma^2}+i\frac{p_0}{\hbar}-\frac{im{\bar\omega} x}{\hbar\sinh{\bar\omega} T},\\
{\bar \zeta}(x,T)=& \frac{x_0}{2\sigma^2}+i\frac{p_0}{\hbar}+\frac{im{\bar\omega} x}{\hbar\sinh{\bar\omega} T}.
\end{aligned} 
\end{align} 
On the other hand, the solution with the Neumann boundary condition and the initial profile (\ref{initial2}) (with $L\to \infty$) is obtained from Eq.~(\ref{harmonic-n}) as
\begin{align}
\Psi(x,T)
=&\sqrt{\frac{m{\bar\omega}}{2\pi i\hbar \sinh{\bar\omega} T}}\exp\biggl(-\frac{x_0^2}{4\sigma^2}+\frac{im{\bar\omega} x^2\cosh{\bar\omega} T}{2\hbar\sinh{\bar\omega} T}\biggl)  \nonumber \\
&\times \biggl[\frac{1}{2\eta}\biggl\{\frac{1}{2}\sqrt{\frac{\pi}{\eta }}\biggl(\zeta e^{\zeta^2/4\eta }+{\bar \zeta}e^{{\bar \zeta}^2/4\eta }\biggl)+i\zeta F\biggl(-\frac{i\zeta}{2\sqrt{\eta }}\biggl)+i{\bar \zeta} F\biggl(-\frac{i{\bar \zeta}}{2\sqrt{\eta }}\biggl)+2\biggl\} \nonumber \\
&-\biggl(\frac{x_0}{2\sigma^2}+i\frac{p_0}{\hbar}\biggl)^{-1} \biggl\{\frac12\sqrt{\frac{\pi}{\eta }}\biggl(e^{\zeta^2/4\eta }+e^{{\bar \zeta}^2/4\eta }\biggl)+\frac{i}{\sqrt{\eta }}F\biggl(-\frac{i\zeta}{2\sqrt{\eta }}\biggl)+\frac{i}{\sqrt{\eta }}F\biggl(-\frac{i{\bar \zeta}}{2\sqrt{\eta }}\biggl)\biggl\}\biggl].  \label{sol-n-harmonic}
\end{align} 

In the limit to the zero cosmological constant $\Lambda\to 0$ (${\bar\omega}\to 0$), the solutions (\ref{sol-d-harmonic}) and (\ref{sol-n-harmonic}) reduce to
\begin{align}
\Psi(x,T)
=&\frac{1}{2\eta }\sqrt{\frac{m}{2\pi i\hbar T}}e^{-x_0^2/4\sigma^2+imx^2/2\hbar T} \nonumber \\
&\times \biggl\{\frac{1}{2}\sqrt{\frac{\pi}{\eta }}\biggl(\zeta e^{\zeta ^2/4\eta }-{\bar \zeta}e^{{\bar \zeta}^2/4\eta }\biggl)+i\zeta F\biggl(-\frac{i\zeta}{2\sqrt{\eta }}\biggl)-i{\bar \zeta}F\biggl(-\frac{i{\bar \zeta}}{2\sqrt{\eta }}\biggl)\biggl\} \label{sol-d-free}
\end{align} 
and
\begin{align}
\Psi(x,T)
=&\sqrt{\frac{m}{2\pi i\hbar T}}e^{-x_0^2/4\sigma^2+imx^2/2\hbar T} \nonumber \\
&\times \biggl[\frac{1}{2\eta}\biggl\{\frac{1}{2}\sqrt{\frac{\pi}{\eta }}\biggl(\zeta e^{\zeta^2/4\eta }+{\bar \zeta}e^{{\bar \zeta}^2/4\eta }\biggl)+i\zeta F\biggl(-\frac{i\zeta}{2\sqrt{\eta }}\biggl)+i{\bar \zeta} F\biggl(-\frac{i{\bar \zeta}}{2\sqrt{\eta }}\biggl)+2\biggl\} \nonumber \\
&-\biggl(\frac{x_0}{2\sigma^2}+i\frac{p_0}{\hbar}\biggl)^{-1} \biggl\{\frac12\sqrt{\frac{\pi}{\eta }}\biggl(e^{\zeta^2/4\eta }+e^{{\bar \zeta}^2/4\eta }\biggl)+\frac{i}{\sqrt{\eta }}F\biggl(-\frac{i\zeta}{2\sqrt{\eta }}\biggl)+\frac{i}{\sqrt{\eta }}F\biggl(-\frac{i{\bar \zeta}}{2\sqrt{\eta }}\biggl)\biggl\}\biggl],  \label{sol-n-free}
\end{align} 
respectively, where complex functions $\eta $, $\zeta $, and ${\bar \zeta}$ are now
\begin{align}
\begin{aligned}
\eta(T) =& \frac{1}{4\sigma^2}-\frac{im}{2\hbar T},\\
\zeta(x,T) =&\frac{x_0}{2\sigma^2}+i\frac{p_0}{\hbar}-\frac{imx}{\hbar T},\\
{\bar \zeta}(x,T)=& \frac{x_0}{2\sigma^2}+i\frac{p_0}{\hbar}+\frac{imx}{\hbar T}.
\end{aligned} 
\end{align}

Although it is difficult to see analytically the bouncing behavior of $\langle x^2\rangle(T)(=\langle a^3\rangle(T))$ in these solutions, we can evaluate its asymptotic behavior for $T\to \infty$.
For the solution with the Dirichlet boundary condition (\ref{sol-d-harmonic}), changing the coordinate as $s:={\bar\omega} x/\sinh{\bar\omega} T$, we obtain
\begin{align}
\int_0^\infty x^q|\Psi(x,T)|^2\D x=&\frac{m}{8\pi \hbar|\eta|^2}\biggl(\frac{\sinh{\bar\omega} T}{{\bar\omega}}\biggl)^qe^{-x_0^2/4\sigma^2} \int_0^\infty s^q \nonumber \\
&\times\biggl\{\frac{1}{2}\sqrt{\frac{\pi}{\eta }}\biggl(\zeta e^{\zeta^2/4\eta }-{\bar \zeta}e^{{\bar \zeta}^2/4\eta }\biggl)+i\zeta F\biggl(-\frac{i\zeta}{2\sqrt{\eta }}\biggl)-i{\bar \zeta} F\biggl(-\frac{i{\bar \zeta}}{2\sqrt{\eta }}\biggl)\biggl\} \nonumber \\
&\times\biggl\{\frac{1}{2}\sqrt{\frac{\pi}{\eta }}\biggl(\zeta e^{\zeta^2/4\eta }-{\bar \zeta}e^{{\bar \zeta}^2/4\eta }\biggl)+i\zeta F\biggl(-\frac{i\zeta}{2\sqrt{\eta }}\biggl)-i{\bar \zeta} F\biggl(-\frac{i{\bar \zeta}}{2\sqrt{\eta }}\biggl)\biggl\}^*\D s,
\end{align} 
where
\begin{align}
\eta = \frac{1}{4\sigma^2}-\frac{im{\bar\omega} \cosh{\bar\omega} T}{2\hbar\sinh{\bar\omega} T},\quad \zeta =\frac{x_0}{2\sigma^2}+i\frac{p_0}{\hbar}-\frac{im}{\hbar}s,\quad {\bar \zeta}= \frac{x_0}{2\sigma^2}+i\frac{p_0}{\hbar}+\frac{im}{\hbar}s.
\end{align} 
Since $\int_0^\infty |\Psi(x,T)|^2\D x$ is constant, we evaluate
\begin{align}
\lim_{T\to \infty}\langle x^q\rangle= \lim_{T\to \infty}\frac{\int_{0}^\infty {x}^q|\Psi({x},T)|^2\D {x}}{\int_{0}^\infty |\Psi({x},T)|^2\D {x}} \propto &\biggl(\frac{\sinh{\bar\omega} T}{{\bar\omega}}\biggl)^q \label{a3-asymp-D}
\end{align}
under the assumption that the limit commutes with the integral. 
We obtain the same result for the solution with the Neumann boundary condition (\ref{sol-n-harmonic}).
In the limit of $\Lambda\to 0$ (${\bar\omega}\to 0$), we obtain
\begin{align}
\lim_{T\to \infty}\langle x^q\rangle \propto T^q
\end{align}
in both cases.

Therefore, the late-time evolution of $\langle a^3\rangle$ is
\begin{align}
\lim_{T\to \infty}\langle a^3 \rangle\propto & \left\{
\begin{array}{ll}
\displaystyle{T^2} &\mbox{for} \quad \Lambda=0,\\
\displaystyle{e^{2{\bar \omega}T}} & \mbox{for}\quad \Lambda>0.
\end{array} \right. 
\end{align}
Since we have ${\bar \omega}=\sqrt{3\Lambda/4}$ in our system, this again shows the convergence to the classical evolution.
Unfortunately, the Feynman kernel is not available to construct solutions with the Robin boundary condition.
We will study such solutions numerically in the next subsection.

\subsection{Numerical results}

In the previous subsection, we studied exact solutions satisfying the Dirichlet or Neumann boundary condition at the origin. However, dynamical properties of solutions with the Robin boundary condition are still not clear.
Here we study this problem by solving the Schr\"odinger equation (\ref{Schro-eq}) with $k=0$ numerically from the initial profile (\ref{initial1}) or (\ref{initial2}).
We treat $L$ as a parameter controlling the boundary condition with fixed values of $\sigma$, $x_0$, and $p_0$ and adopt the Planck unit $G=\hbar=c=1$.
In addition, we set $V_0=1$ which means that the spatial volume of the universe is the Planck volume when $a$ (and hence $x$) is unity.

In our numerical calculations, we set the space step $\Delta x=0.05$ and time step $\Delta T=0.05^2/4$.
We confirmed that even with the smaller stepsize the results in the figures and tables are unchanged. 
We also confirmed that the exact solution (\ref{exact-I}) with $n=1$ and $2$ can be constructed numerically in the period of time shown in Fig.~\ref{wave-exact}.
In the long-time computations, however, we could not keep enough accuracy to verify the convergence of $\langle a^3\rangle$ to the classical evolution. 

\begin{figure}[htbp]
\begin{center}
\includegraphics[width=0.85\linewidth]{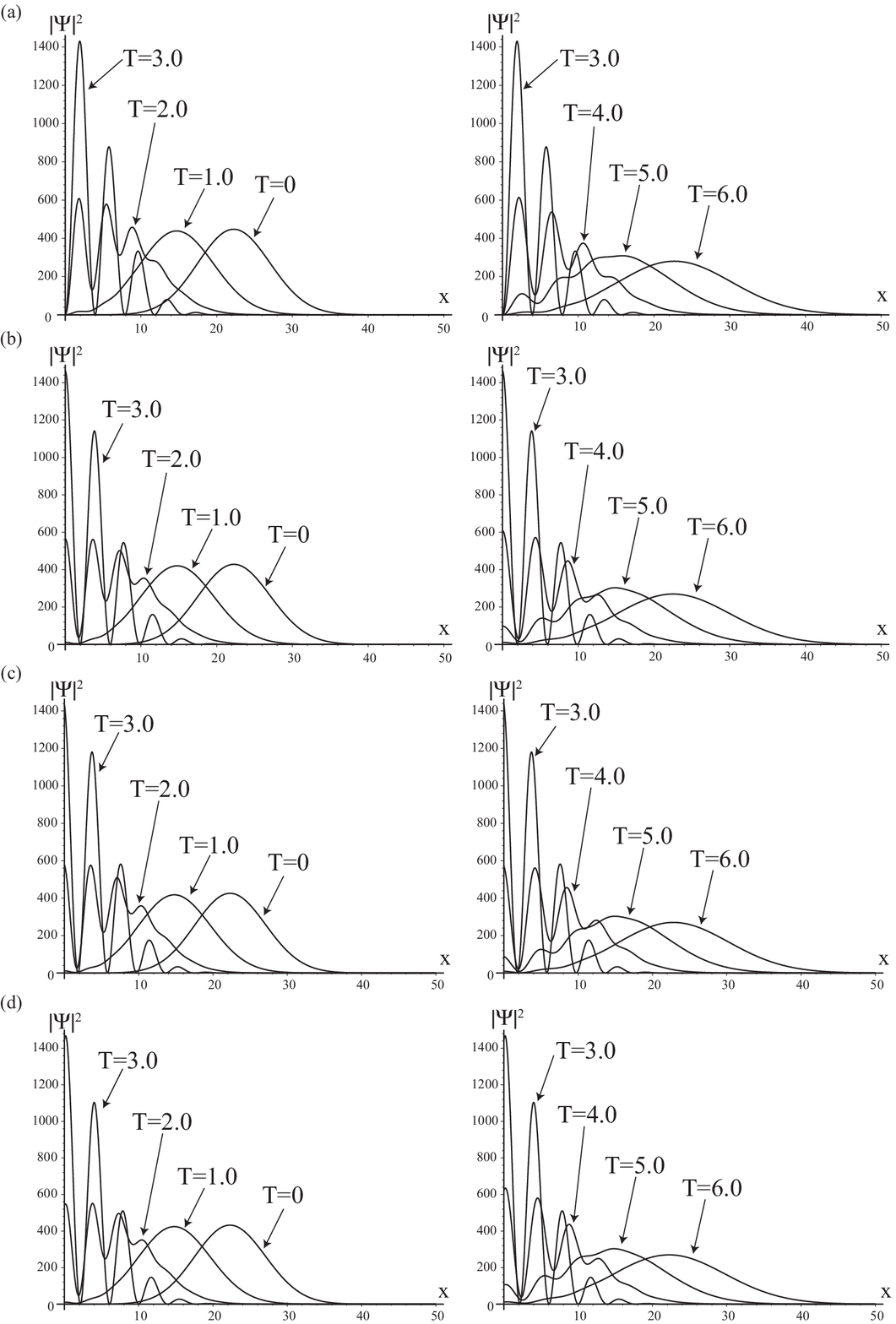}
\caption{\label{wave1} Time evolution of $|\Psi|^2$ before (left) and after (right) the big bounce (around $T=3.0$) for (a) $L=0$, (b) $L=\pm\infty$, (c) $L=10$, and (d) $L=-10$ with $\Lambda=0$ and $x_0=20$, $\sigma=5$, and $p_0=0.8$.
}
\end{center}
\end{figure}

\begin{figure}[htbp]
\begin{center}
\includegraphics[width=0.85\linewidth]{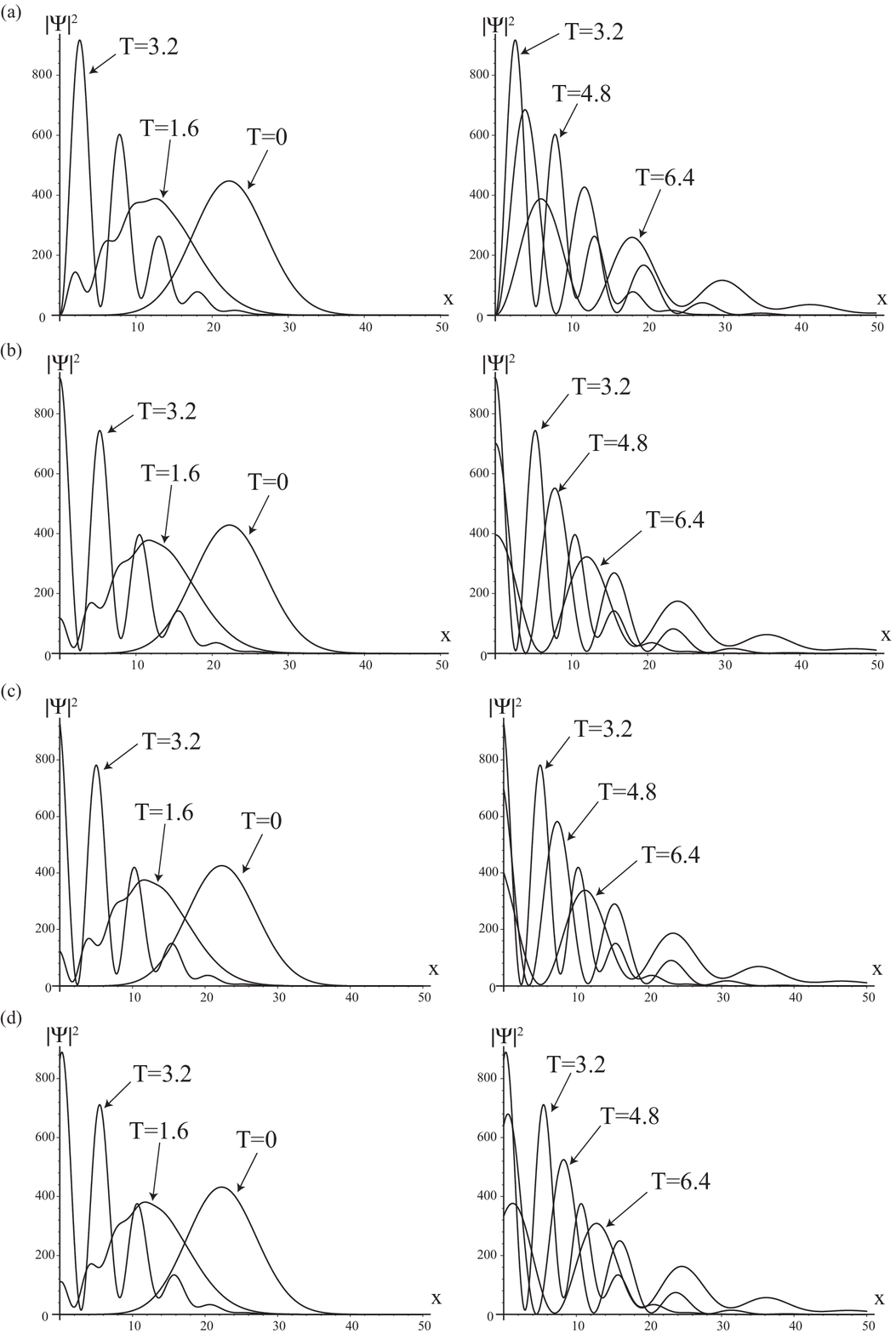}
\caption{\label{wave2} Time evolution of $|\Psi|^2$ before (left) and after (right) the big bounce (around $T=3.2$) for (a) $L=0$, (b) $L=\pm\infty$, (c) $L=10$, and (d) $L=-10$ with $\Lambda=0.1$ and $x_0=20$, $\sigma=5$, and $p_0=0.8$.
}
\end{center}
\end{figure}

\begin{figure}[htbp]
\begin{center}
\includegraphics[width=1.0\linewidth]{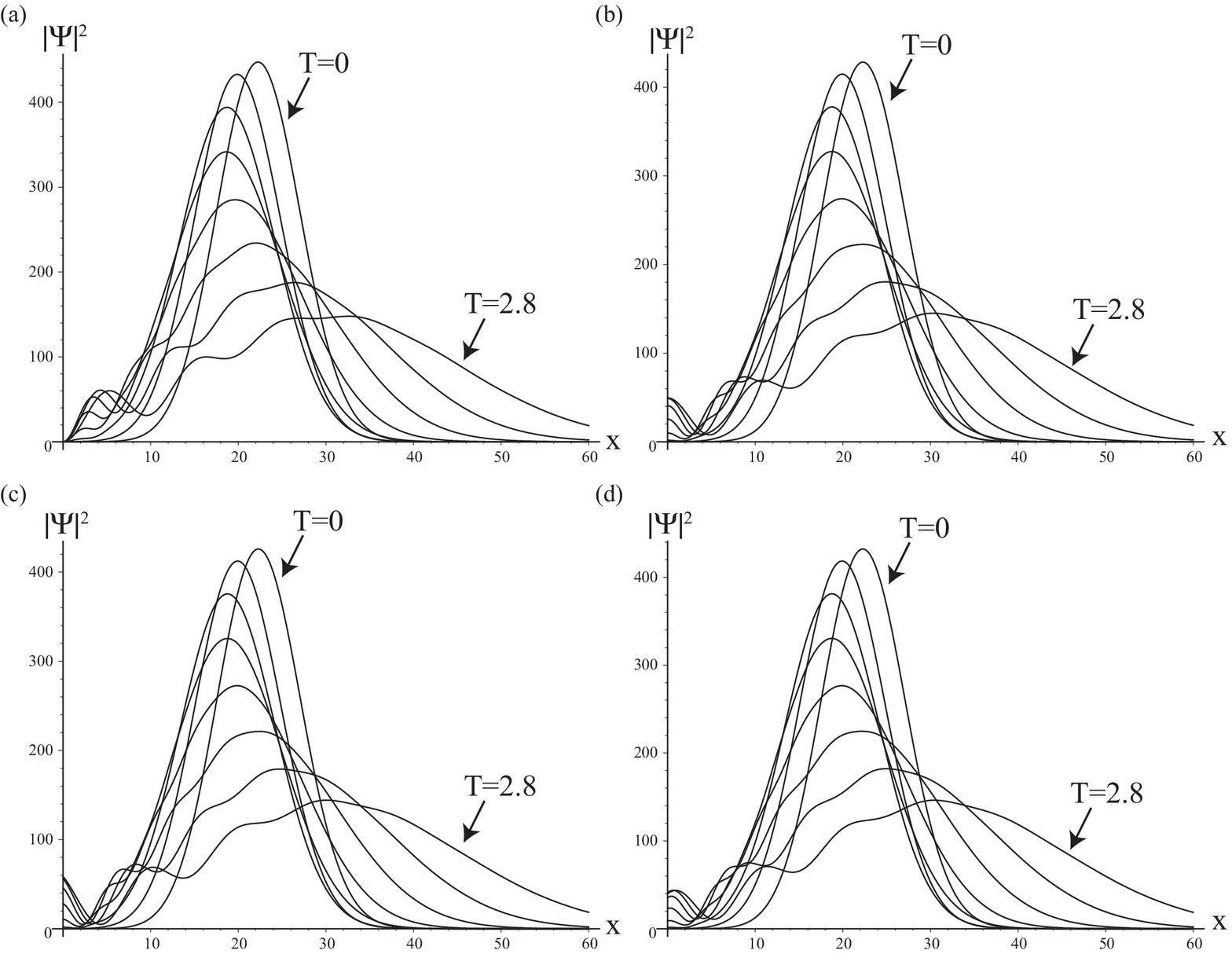}
\caption{\label{wave3} Time evolution of $|\Psi|^2$ around the big bounce (around $T=1.2$) for (a) $L=0$, (b) $L=\pm\infty$, (c) $L=10$, and (d) $L=-10$ with $\Lambda=0.5$ and $x_0=20$, $\sigma=5$, and $p_0=0.8$.
}
\end{center}
\end{figure}

\begin{figure}[htbp]
\begin{center}
\includegraphics[width=1.0\linewidth]{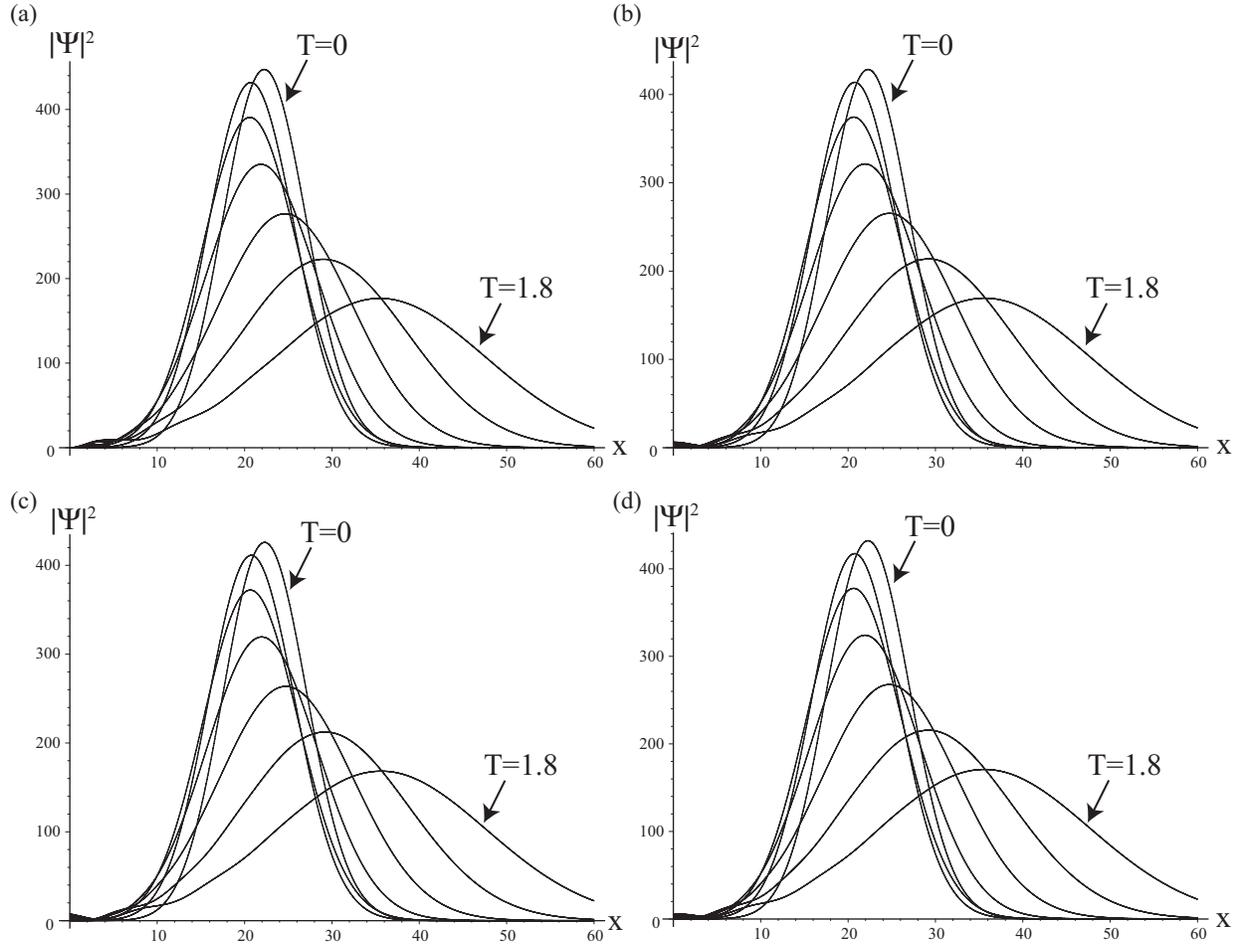}
\caption{\label{wave4} Time evolution of $|\Psi|^2$around the big bounce (around $T=0.6$) for (a) $L=0$, (b) $L=\pm\infty$, (c) $L=10$, and (d) $L=-10$ with $\Lambda=1.0$ and $x_0=20$, $\sigma=5$, and $p_0=0.8$.
}
\end{center}
\end{figure}

We solved the Schr\"odinger equation (\ref{Schro-eq}) numerically by Maple with $x_0=20$, $p_0=0.8$, and $\sigma=5$.
The time evolutions of $|\Psi|^2(x,T)$ with $\Lambda=0$, $0.1$, $0.5$ and $1.0$ are shown in Figs~\ref{wave1}--\ref{wave4}, respectively.
The wave function is regular everywhere during the evolution and it contracts initially and moves back after some moment.
The big bounce behaviors of $\langle a^3\rangle(=\langle x^2\rangle)$ are shown in Tables~\ref{table:shorttime1}--\ref{table:shorttime4}.
For a fixed value of $\Lambda$, the difference in the profiles of $|\Psi|^2$ with different values of $L$ appears only around the origin.
As $\Lambda$ increases, the big bounce occurs sooner and there appears less oscillation around then.
\begin{table}[h]
\caption{\label{table:shorttime1} Values of $\langle x^2\rangle(=\langle a^3\rangle)$ for various values of $L$ in the case of $\Lambda=0$ with $x_0=20$, $\sigma=5$, and $p_0=0.8$.
}
\begin{center}
\begin{tabular}{l@{\qquad}|c@{\qquad}c@{\qquad}c@{\qquad}c@{\qquad}c@{\qquad}c}
\hline \hline
 $T$  &$L=0$ & $L=\pm\infty$ & $L=10$ & $L=-10$ &$L=20$ &$L=-20$　\\\hline
$0$ & { 522.06} &  { 524.08} &  {  524.41} &  {  523.62} &  {  524.26} &  {  523.86}  \\
$0.5$ & {  368.02} &  { 370.06} &  {  370.30} &  {  369.72} &  {  370.19} &  {  369.90}  \\ 
$1.0$ & {  242.89} &  { 244.78} &  {  244.94} &  {  244.53} &  {  244.87} &  {  244.67}  \\ 
$1.5$ & {  146.66} &  { 148.23} &  {  148.33} &  {  148.06} &  {  148.29} &  {  148.15}  \\ 
$2.0$ & {  79.332} &  { 80.407} &  {  80.550} &  {  80.232} &  {  80.482} &  {  80.323}  \\ 
$2.5$ & {  40.910} &  { 41.319} &  {  41.777} &  {  40.880} &  {  41.547} &  {  41.096}  \\ 
$3.0$ & {  31.392} &  { 30.963} &  {  32.273} &  {  29.748} &  {  31.612} &  {  30.337}  \\ 
$3.5$ & {  50.779} &  { 49.338} &  {  52.123} &  {  46.729} &  {  50.724} &  {  47.997}  \\ 
$4.0$ & {  99.069} &  { 96.446} &  {  101.19} &  {  91.949} &  {  98.814} &  {  94.140}  \\ 
$4.5$ & {  176.26} &  { 172.29} &  {  179.26} &  {  165.61} &  {  175.78} &  {  168.87}  \\ 
$5.0$ & {  282.36} &  { 276.86} &  {  286.19} &  {  267.86} &  {  281.54} &  {  272.26}  \\ 
$5.5$ & {  417.36} &  { 410.16} &  {  421.92} &  {  398.78} &  {  416.07} &  {  404.35}  \\ 
\hline
\hline
\end{tabular}
\end{center}
\end{table} 
\begin{table}[h]
\caption{\label{table:shorttime2} Values of $\langle x^2\rangle(=\langle a^3\rangle)$ for various values of $L$ in the case of $\Lambda=0.1$ with $x_0=20$, $\sigma=5$, and $p_0=0.8$.
}
\begin{center}
\begin{tabular}{l@{\qquad}|c@{\qquad}c@{\qquad}c@{\qquad}c@{\qquad}c@{\qquad}c}
\hline \hline
 $T$  &$L=0$ & $L=\pm\infty$ & $L=10$ & $L=-10$ &$L=20$ &$L=-20$　\\\hline
$0$ & { 522.06} &  { 524.08} &  {  524.41} &  {  523.62} &  {  524.26} &  {  523.86}  \\
$0.5$ & {  375.84} &  { 377.92} &  {  378.16} &  {  378.06} &  {  377.57} &  {  377.76}  \\ 
$1.0$ & {  267.37} &  { 269.42} &  {  269.59} &  {  269.52} &  {  269.15} &  {  269.30}  \\ 
$1.5$ & {  188.46} &  { 190.38} &  {  190.52} &  {  190.46} &  {  190.17} &  {  190.29}  \\ 
$2.0$ & {  133.17} &  { 134.86} &  {  135.01} &  {  134.94} &  {  134.65} &  {  134.76}  \\ 
$2.5$ & {  97.318} &  { 98.658} &  {  98.991} &  {  98.828} &  {  98.304} &  {  98.483}  \\ 
$3.0$ & {  78.206} &  { 79.045} &  {  79.877} &  {  79.458} &  {  78.263} &  {  78.645}  \\ 
$3.5$ & {  74.390} &  { 74.543} &  {  76.325} &  {  75.424} &  {  72.923} &  {  73.703}  \\ 
$4.0$ & {  85.582} &  { 84.810} &  {  88.095} &  {  86.435} &  {  81.853} &  {  83.270}  \\ 
$4.5$ & {  112.63} &  { 110.62} &  {  116.05} &  {  113.31} &  {  105.75} &  {  108.08}  \\ 
$5.0$ & {  157.56} &  { 153.92} &  {  162.24} &  {  158.05} &  {  146.45} &  {  150.02}  \\ 
$5.5$ & {  223.77} &  { 217.97} &  {  230.11} &  {  224.00} &  {  207.09} &  {  212.28}  \\ 
\hline
\hline
\end{tabular}
\end{center}
\end{table} 
\begin{table}[h]
\caption{\label{table:shorttime3} Values of $\langle x^2\rangle(=\langle a^3\rangle)$ for various values of $L$ in the case of $\Lambda=0.5$ with $x_0=20$, $\sigma=5$, and $p_0=0.8$.
}
\begin{center}
\begin{tabular}{l@{\qquad}|c@{\qquad}c@{\qquad}c@{\qquad}c@{\qquad}c@{\qquad}c}
\hline \hline
 $T$  &$L=0$ & $L=\pm\infty$ & $L=10$ & $L=-10$ &$L=20$ &$L=-20$　\\\hline
$0$ & { 522.06} &  { 524.08} &  {  524.41} &  {  523.62} &  {  524.26} &  {  523.86}  \\
$0.2$ & {  464.18} &  { 466.26} &  {  466.55} &  {  466.42} &  {  465.84} &  {  466.06}  \\ 
$0.4$ & {  423.19} &  { 425.36} &  {  425.63} &  {  425.51} &  {  424.97} &  {  425.18}  \\ 
$0.6$ & {  396.62} &  { 398.93} &  {  399.19} &  {  399.08} &  {  398.56} &  {  398.76}  \\ 
$0.8$ & {  382.88} &  { 385.38} &  {  385.63} &  {  385.52} &  {  385.01} &  {  385.21}  \\ 
$1.0$ & {  381.13} &  { 383.89} &  {  384.14} &  {  384.03} &  {  383.51} &  {  383.72}  \\ 
$1.2$ & {  391.27} &  { 394.37} &  {  394.62} &  {  394.51} &  {  393.97} &  {  394.19}  \\ 
$1.4$ & {  413.92} &  { 417.44} &  {  417.72} &  {  417.60} &  {  417.01} &  {  417.25}  \\ 
$1.6$ & {  450.43} &  { 454.50} &  {  454.81} &  {  454.67} &  {  454.02} &  {  454.28}  \\ 
$1.8$ & {  502.99} &  { 507.78} &  {  508.12} &  {  507.97} &  {  507.23} &  {  507.53}  \\ 
$2.0$ & {  574.77} &  { 580.47} &  {  580.87} &  {  580.70} &  {  579.83} &  {  580.18}  \\ 
$2.2$ & {  670.08} &  { 676.94} &  {  677.42} &  {  677.22} &  {  676.18} &  {  676.60}  \\ 
\hline
\hline
\end{tabular}
\end{center}
\end{table} 
\begin{table}[h]
\caption{\label{table:shorttime4} Values of $\langle x^2\rangle(=\langle a^3\rangle)$ for various values of $L$ in the case of $\Lambda=1.0$ with $x_0=20$, $\sigma=5$, and $p_0=0.8$.
}
\begin{center}
\begin{tabular}{l@{\qquad}|c@{\qquad}c@{\qquad}c@{\qquad}c@{\qquad}c@{\qquad}c}
\hline \hline
 $T$  &$L=0$ & $L=\pm\infty$ & $L=10$ & $L=-10$ &$L=20$ &$L=-20$　\\\hline
$0$ & { 522.06} &  { 524.08} &  {  524.41} &  {  523.62} &  {  524.26} &  {  523.86}  \\
$0.1$ & {  492.70} &  { 494.75} &  {  495.06} &  {  494.31} &  {  494.92} &  {  494.55}  \\ 
$0.2$ & {  471.45} &  { 473.56} &  {  473.87} &  {  473.14} &  {  473.73} &  {  473.37}  \\ 
$0.3$ & {  457.70} &  { 459.89} &  {  460.19} &  {  459.48} &  {  460.05} &  {  459.70}  \\ 
$0.4$ & {  451.02} &  { 453.32} &  {  453.61} &  {  452.91} &  {  453.48} &  {  453.13}  \\ 
$0.5$ & {  451.20} &  { 453.66} &  {  453.95} &  {  453.24} &  {  453.82} &  {  453.46}  \\ 
$0.6$ & {  458.27} &  { 460.91} &  {  461.20} &  {  460.48} &  {  461.07} &  {  460.71}  \\ 
$0.7$ & {  472.43} &  { 475.29} &  {  475.59} &  {  474.84} &  {  475.46} &  {  475.08}  \\ 
$0.8$ & {  494.10} &  { 497.23} &  {  497.55} &  {  496.75} &  {  497.41} &  {  497.01}  \\ 
$0.9$ & {  523.93} &  { 527.39} &  {  527.73} &  {  526.88} &  {  527.58} &  {  527.16}  \\ 
$1.0$ & {  562.82} &  { 566.67} &  {  567.04} &  {  566.12} &  {  566.87} &  {  566.42}  \\ 
$1.1$ & {  611.92} &  { 616.25} &  {  616.64} &  {  615.64} &  {  616.47} &  {  615.97}  \\ 
\hline
\hline
\end{tabular}
\end{center}
\end{table} 

\section{Summary}
\label{sec:summary}
In the present paper, we have studied the time evolution of a wave function for the spatially flat FLRW universe governed by the Wheeler-DeWitt equation, with or without a positive cosmological constant $\Lambda$.
We have adopted the Laplace-Beltrami operator-ordering and considered a Brown-Kucha\v{r} dust as a matter field.
Then the system has reduced to quantum mechanics on the half-line and the Wheeler-DeWitt equation has the form of the time-dependent Schr\" odinger equation for a harmonic oscillator with negative mass, where a scalar field $T$ acts as a time variable. 

Self-adjoint extension of the Hamiltonian operator admits a one-parameter family of boundary conditions at the origin in the minisuperspace.
For any value of the extension parameter $L$, the time evolution of a wave function is unitary and the corresponding quantum system is totally well-defined.
We have shown that the classical initial singularity is avoided and replaced by the big bounce in the quantum system.
These properties have been shown also under the different operator-orderings in~\cite{ak}.

We have also studied the problem whether the expectation value of the spatial volume of the universe $\langle a^3\rangle$ obeys the classical evolution in the late time.
This is a nontrivial problem because the Ehrenfest's theorem is not valid in quantum mechanics on the half line.
We have used exact solutions with the Dirichlet or Neumann boundary condition at the origin and analytically showed the convergence to the classical evolution both in the cases with and without $\Lambda$.
However, this does not mean the classicalization of the quantum universe in the late time because the variance of $a^3$ is diverging.

Unfortunately, exact solutions are not available in the case of the Robin boundary condition.
In order to verify the convergence in such cases, long-time numerical computations with high accuracy are required.
Such numerical studies will also clarify the effect of the spatial curvature of the universe on the evolution.
These are left for future investigations.

Another promising direction of future research is to generalize our study in a more general minisuperspace such as the Bianchi minisuperspace.
In such cases, the corresponding quantum systems are higher-dimensional and the problems of the self-adjoint extension and initial-singularity avoidance are highly nontrivial.
Also, quantization of the inhomogeneous universe is another possible direction.
Classically, the general spherically symmetric solution with a dust fluid is the Lema{\^ i}tre-Tolman-Bondi solution.
The corresponding quantum system is infinite dimensional, namely a quantum field theory, and therefore a totally different treatment of the system is needed~\cite{midi-review}.
Those results could shed light on generic properties of canonical quantum cosmology and will be reported elsewhere.

\subsection*{Acknowledgments}
The author thanks Jorge Zanelli, Francisco Correa, and Gabor Kunstatter for valuable comments. 
The author also thanks the anonymous referees for their careful reading of the manuscript and valuable comments, which significantly contributed to improving the quality of the publication.
The author is grateful to Tatsuhiko Koike for discussions at the initial stage of the present work.

\appendix

\section{Six-parameter family of exact solutions}
\label{app:six}
For the following Schr\" odinger equation for a harmonic oscillator with positive mass;
\begin{align}
-\frac12\frac{\partial^2 \Psi}{\partial x^2}+\frac12k^2x^2\Psi=i\frac{\partial \Psi}{\partial t},
\end{align}
there is a six-parameter family of exact solutions~\cite{lsv2013}\footnote{We have done coordinate transformations and reparametrization from the original expressions in~\cite{lsv2013}.}:
\begin{align}
\Psi(x,t)=\Psi_n(x,t):=&\frac{e^{i\left(\alpha(t)x^2+\delta(t)x+\kappa(t)\right)+i(2n+1)\gamma(t)}}{\sqrt{2^nn!\mu(t)\sqrt{\pi}}}e^{-(\beta(t)x+\varepsilon(t))^2/2}H_n(\beta(t)x+\varepsilon(t)), \label{six-solution}
\end{align}
where $H_n(x)$ is the Hermite polynomials and 
\begin{align}
\mu(t)=&\mu_0\sqrt{{\bar\beta}_0^4\sin^2kt/k^2+(2\alpha_0\sin kt+\cos kt)^2}, \label{six-sol1}\\
\alpha(t)=&\frac{k\alpha_0\cos 2kt+(\sin 2kt/k)({\bar\beta}_0^4+4k^2\alpha_0^2-k^2)/4}{{\bar\beta}_0^4\sin^2kt/k^2+(2\alpha_0\sin kt+\cos kt)^2},\\
\beta(t)=&\frac{{\bar\beta}_0}{\sqrt{{\bar\beta}_0^4\sin^2kt/k^2+(2\alpha_0\sin kt+\cos kt)^2}},\\
\gamma(t)=&\gamma_0-\frac12\arctan\biggl(\frac{{\bar\beta}_0^2\sin kt/k}{2\alpha_0\sin kt+\cos kt}\biggl),\\
\delta(t)=&\frac{{\bar\delta}_0(2\alpha_0\sin kt+\cos kt)+\varepsilon_0{\bar\beta}_0^3\sin kt/k}{{\bar\beta}_0^4\sin^2kt/k^2+(2\alpha_0\sin kt+\cos kt)^2},\\
\varepsilon(t)=&\frac{\varepsilon_0(2\alpha_0\sin kt+\cos kt)-{\bar\beta}_0{\bar\delta}_0\sin kt/k}{\sqrt{{\bar\beta}_0^4\sin^2kt/k^2+(2\alpha_0\sin kt+\cos kt)^2}},\\
\kappa(t)=&\kappa_0+\frac{\sin^2 kt}{k^2}\frac{\varepsilon_0{\bar\beta}_0^2(k\alpha_0\varepsilon_0-{\bar\beta}_0{\bar\delta}_0)-k\alpha_0{\bar\delta}_0^2}{{\bar\beta}_0^4\sin^2kt/k^2+(2\alpha_0\sin kt+\cos kt)^2} \nonumber \\
&+\frac14\frac{\sin 2kt}{k}\frac{\varepsilon_0^2{\bar\beta}_0^2-{\bar\delta}_0^2}{{\bar\beta}_0^4\sin^2kt/k^2+(2\alpha_0\sin kt+\cos kt)^2}.\label{six-sol2}
\end{align}
One of the seven parameters $\mu_0,{\alpha}_0,{\bar \beta}_0,\gamma_0,{\bar\delta}_0,\kappa_0,\varepsilon_0$ may be used for normalization, so that the number of independent parameters is six.
The norm of $\Psi_n$ in the case of the full line is given by 
\begin{align}
\int_{-\infty}^\infty|\Psi_n(x,t)|^2\D x=\frac{1}{\mu_0{\bar \beta}_0},
\end{align}
where we used
\begin{align}
\int_{-\infty}^\infty H_m(x)H_n(x)e^{-x^2}\D x=\sqrt{\pi}2^nn!\delta_{mn}.
\end{align}

The solution for a free particle is obtained in the limit $k\to 0$ as
\begin{align}
\mu(t)=&\mu_0\sqrt{{\bar\beta}_0^4t^2+1},\qquad \alpha(t)=\frac{t{\bar\beta}_0^4/2}{{\bar\beta}_0^4t^2+1},\\
\beta(t)=&\frac{{\bar\beta}_0}{\sqrt{{\bar\beta}_0^4t^2+1}},\qquad \gamma(t)=\gamma_0-\frac12\arctan({\bar\beta}_0^2t),\\
\delta(t)=&\frac{{\bar\delta}_0+\varepsilon_0{\bar\beta}_0^3t}{{\bar\beta}_0^4t^2+1},\qquad \varepsilon(t)=\frac{\varepsilon_0-{\bar\beta}_0{\bar\delta}_0t}{\sqrt{{\bar\beta}_0^4t^2+1}},\\
\kappa(t)=&\kappa_0-t^2\frac{\varepsilon_0{\bar\beta}_0^3{\bar\delta}_0}{{\bar\beta}_0^4t^2+1} +\frac12 t\frac{\varepsilon_0^2{\bar\beta}_0^2-{\bar\delta}_0^2}{{\bar\beta}_0^4t^2+1},
\end{align}
where the parameter $\alpha_0$ has disappeared.
By the reparametrization $k=i{\bar k}$ and $i\alpha_0={\bar\alpha}_0$ in Eqs.~(\ref{six-sol1})--(\ref{six-sol2}), we obtain the solution for an inverted-harmonic oscillator as
\begin{align}
\mu(t)=&\mu_0\sqrt{{\bar\beta}_0^4\sinh^2{\bar k}t/{\bar k}^2+(2{\bar\alpha}_0\sinh{\bar k}t+\cosh{\bar k}t)^2},\label{six-anti1} \\
\alpha(t)=&\frac{{\bar k}{\bar\alpha}_0\cosh 2{\bar k}t+(\sinh 2{\bar k}t/{\bar k})({\bar\beta}_0^4-4{\bar k}^2\alpha_0^2+{\bar k}^2)/4}{{\bar\beta}_0^4\sinh^2{\bar k}t/{\bar k}^2+(2{\bar\alpha}_0\sinh{\bar k}t+\cosh{\bar k}t)^2},\\
\beta(t)=&\frac{{\bar\beta}_0}{\sqrt{{\bar\beta}_0^4\sinh^2{\bar k}t/{\bar k}^2+(2{\bar\alpha}_0\sinh{\bar k}t+\cosh{\bar k}t)^2}},\\
\gamma(t)=&\gamma_0-\frac12\arctan\biggl(\frac{{\bar\beta}_0^2\sinh {\bar k}t/{\bar k}}{2{\bar\alpha}_0\sinh{\bar k}t+\cosh{\bar k}t}\biggl),\\
\delta(t)=&\frac{{\bar\delta}_0(2{\bar\alpha}_0\sinh{\bar k}t+\cosh{\bar k}t)+\varepsilon_0{\bar\beta}_0^3\sinh {\bar k}t/{\bar k}}{{\bar\beta}_0^4\sinh^2{\bar k}t/{\bar k}^2+(2{\bar\alpha}_0\sinh{\bar k}t+\cosh{\bar k}t)^2},\\
\varepsilon(t)=&\frac{\varepsilon_0(2{\bar\alpha}_0\sinh{\bar k}t+\cosh{\bar k}t)-{\bar\beta}_0{\bar\delta}_0\sinh {\bar k}t/{\bar k}}{\sqrt{{\bar\beta}_0^4\sinh^2{\bar k}t/{\bar k}^2+(2{\bar\alpha}_0\sinh{\bar k}t+\cosh{\bar k}t)^2}},\\
\kappa(t)=&\kappa_0+\frac{\sinh^2 {\bar k}t}{{\bar k}^2}\frac{\varepsilon_0{\bar\beta}_0^2({\bar k}{\bar \alpha}_0\varepsilon_0-{\bar\beta}_0{\bar\delta}_0)-{\bar k}{\bar \alpha}_0{\bar\delta}_0^2}{{\bar\beta}_0^4\sinh^2{\bar k}t/{\bar k}^2+(2{\bar\alpha}_0\sinh{\bar k}t+\cosh{\bar k}t)^2} \nonumber \\
&+\frac14\frac{\sinh 2{\bar k}t}{{\bar k}}\frac{\varepsilon_0^2{\bar\beta}_0^2-{\bar\delta}_0^2}{{\bar\beta}_0^4\sinh^2{\bar k}t/{\bar k}^2+(2{\bar\alpha}_0\sinh{\bar k}t+\cosh{\bar k}t)^2}.\label{six-anti2}
\end{align}
The solution of the Schr\" odinger equation~(\ref{schr1}) for a harmonic oscillator with negative mass is obtained from the above by the transformation $t\to -T$.

\section{Exact solutions constructed with the Feynman kernel}
\label{app:kernel}
In this appendix, we present exact time-dependent wave functions on the half-line constructed with the Feynman kernel.
In the case of the full line, the time-dependent solution of the Schr\"odinger equation
\begin{align}
-\frac{\hbar^2}{2m}\frac{\partial^2\Phi}{\partial x^2}+V(x)\Phi=i\hbar \frac{\partial \Phi}{\partial T} \label{Schro}
\end{align}  
with an initial profile $\Phi(x,0)$ is given by
\begin{align}
\Phi(x,t)=\int_{-\infty}^\infty K(x,t;x',0)\Phi(x',0)\D x'.
\end{align}  
For a free particle or harmonic oscillator, the Feynman kernel $K(x,t;x',0)$ is known~\cite{Sakurai}.
In the case of the harmonic oscillator $V(x)=(1/2)mw^2x^2$, it is given by
\begin{align}
K(x,t;x',0)=&\sqrt{\frac{m\omega}{2\pi i\hbar \sin\omega t}}\exp\biggl(\frac{im\omega\{(x^2+{x'}^2)\cos\omega t-2xx'\}}{2\hbar\sin\omega t}\biggl).
\end{align}  
Taking the limit $\omega\to 0$, we obtain the Feynman kernel for a free particle:
\begin{align}
K(x,t;x',0)=&\sqrt{\frac{m}{2\pi i\hbar t}}\exp\biggl(\frac{im(x-{x'})^2}{2\hbar t}\biggl),
\end{align}  
which is also written as
\begin{align}
K(x,t;x',0)=&\frac{1}{2\pi}\int_{-\infty}^\infty e^{ik(x-x')-i\hbar tk^2/2m}\D k.
\end{align}

Using them, we can construct time-dependent solutions on the half-line with the Dirichlet or Neumann boundary condition at $x=0$.
The solution is then given by
\begin{align}
\Phi(x,t)=\int_0^\infty K(x,t;x',0)\Phi(x',0)\D x' \label{time-sol}
\end{align}  
and the expressions of $K(x,t;x',0)$ will be shown below.
For derivation, we will use the following Formula 7.7.3 in~\cite{NIST}:
\begin{align}
\int_{0}^\infty e^{-\eta {x'}^2+2izx'}\D x'=\frac12\sqrt{\frac{\pi}{\eta }}e^{-z^2/\eta }+\frac{i}{\sqrt{\eta }}F(z/\sqrt{\eta })~~~({\rm Re}(\eta) >0),
\end{align} 
where $F(z)$ is the Dawson function defined by 
\begin{align}
F(z):=e^{-z^2}\int_0^ze^{w^2}\D w.
\end{align} 
We will also use the following:
\begin{align}
\int_{0}^\infty x' e^{-\eta {x'}^2+2izx'}\D x'=&\frac{1}{2i}\frac{\D }{\D z}\biggl(\int_{0}^\infty e^{-\eta {x'}^2+2izx'}\D x'\biggl) \nonumber \\
=&\frac{1}{2\eta}\biggl(iz\sqrt{\frac{\pi}{\eta }}e^{-z^2/\eta }-2zF(z/\sqrt{\eta })+1\biggl),
\end{align} 
where we used $\D F/\D z=-2zF+1$.

\subsection{Harmonic oscillator}

In the case of the harmonic oscillator on the half-line, the Feynman kernel is given by
\begin{align}
K(x,t;x',0)=&\sqrt{\frac{m\omega}{2\pi i\hbar \sin\omega t}}\biggl\{\exp\biggl(\frac{im\omega\{(x^2+{x'}^2)\cos\omega t-2xx'\}}{2\hbar\sin\omega t}\biggl) \nonumber \\
&\mp \exp\biggl(\frac{im\omega\{(x^2+{x'}^2)\cos\omega t+2xx'\}}{2\hbar\sin\omega t}\biggl)\biggl\},
\end{align}  
where the minus (plus) sign corresponds to the Dirichlet (Neumann) boundary condition.
From our initial profile with the Dirichlet boundary condition;
\begin{align}
\Phi(x,0)=x\exp\biggl(-\frac{(x-x_0)^2}{4\sigma^2}+i\frac{p_0x}{\hbar}\biggl),
\end{align}  
the time-dependent solution is  
\begin{align}
\Phi(x,t)
=&\frac{1}{2\eta}\sqrt{\frac{m\omega}{2\pi i\hbar \sin\omega t}}\exp\biggl(-\frac{x_0^2}{4\sigma^2}+\frac{im\omega x^2\cos\omega t}{2\hbar\sin\omega t}\biggl)  \nonumber \\
&\times\biggl\{\frac{1}{2}\sqrt{\frac{\pi}{\eta }}\biggl(\zeta e^{\zeta^2/4\eta }-{\bar \zeta}e^{{\bar \zeta}^2/4\eta }\biggl)+i\zeta F\biggl(-\frac{i\zeta}{2\sqrt{\eta }}\biggl)-i{\bar \zeta} F\biggl(-\frac{i{\bar \zeta}}{2\sqrt{\eta }}\biggl)\biggl\}, \label{harmonic-d}
\end{align} 
where complex functions $\eta $, $\zeta $, and ${\bar \zeta}$ are defined by
\begin{align}
\begin{aligned}
\eta(t) :=& \frac{1}{4\sigma^2}-\frac{im\omega \cos\omega t}{2\hbar\sin\omega t},\\
\zeta(x,t) :=&\frac{x_0}{2\sigma^2}+i\frac{p_0}{\hbar}-\frac{im\omega x}{\hbar\sin\omega t},\\
{\bar \zeta}(x,t):=& \frac{x_0}{2\sigma^2}+i\frac{p_0}{\hbar}+\frac{im\omega x}{\hbar\sin\omega t}.
\end{aligned} 
\end{align} 

On the other hand, from our initial profile with the Neumann boundary condition;
\begin{align}
\Phi(x,0)=&\biggl\{x-\biggl(\frac{x_0}{2\sigma^2}+i\frac{p_0}{\hbar}\biggl)^{-1}\biggl\}\exp\biggl(-\frac{(x-x_0)^2}{4\sigma^2}+i\frac{p_0x}{\hbar}\biggl),
\end{align}  
the time-dependent solution is 
\begin{align}
\Phi(x,t)
=&\sqrt{\frac{m\omega}{2\pi i\hbar \sin\omega t}}\exp\biggl(-\frac{x_0^2}{4\sigma^2}+\frac{im\omega x^2\cos\omega t}{2\hbar\sin\omega t}\biggl)  \nonumber \\
&\times \biggl[\frac{1}{2\eta}\biggl\{\frac{1}{2}\sqrt{\frac{\pi}{\eta }}\biggl(\zeta e^{\zeta^2/4\eta }+{\bar \zeta}e^{{\bar \zeta}^2/4\eta }\biggl)+i\zeta F\biggl(-\frac{i\zeta}{2\sqrt{\eta }}\biggl)+i{\bar \zeta} F\biggl(-\frac{i{\bar \zeta}}{2\sqrt{\eta }}\biggl)+2\biggl\} \nonumber \\
&-\biggl(\frac{x_0}{2\sigma^2}+i\frac{p_0}{\hbar}\biggl)^{-1} \biggl\{\frac12\sqrt{\frac{\pi}{\eta }}\biggl(e^{\zeta^2/4\eta }+e^{{\bar \zeta}^2/4\eta }\biggl)+\frac{i}{\sqrt{\eta }}F\biggl(-\frac{i\zeta}{2\sqrt{\eta }}\biggl)+\frac{i}{\sqrt{\eta }}F\biggl(-\frac{i{\bar \zeta}}{2\sqrt{\eta }}\biggl)\biggl\}\biggl]. \label{harmonic-n}
\end{align}

\subsection{Free particle}
The expressions for a free particle ($V(x)=0$) are obtained in the limit $\omega\to 0$ from the ones in the previous subsection.
The Feynman kernel is given by
\begin{align}
K(x,t;x',0)=&\sqrt{\frac{m}{2\pi i\hbar t}}\biggl(e^{im(x-x')^2/2\hbar t} \mp e^{im(x+x')^2/2\hbar t}\biggl) \nonumber \\
=&\frac{1}{2\pi}\int_{-\infty}^\infty e^{-i\hbar tk^2/2m}\biggl(e^{ik(x-x')} \mp e^{ik(x+x')}\biggl)\D k.
\end{align} 
The time-dependent solution from our initial profile with the Dirichlet boundary condition is given as
\begin{align}
\Phi(x,t)
=&\frac{1}{2\eta }\sqrt{\frac{m}{2\pi i\hbar t}}e^{-x_0^2/4\sigma^2+imx^2/2\hbar t} \nonumber \\
&\times \biggl\{\frac{1}{2}\sqrt{\frac{\pi}{\eta }}\biggl(\zeta e^{\zeta ^2/4\eta }-{\bar \zeta}e^{{\bar \zeta}^2/4\eta }\biggl)+i\zeta F\biggl(-\frac{i\zeta}{2\sqrt{\eta }}\biggl)-i{\bar \zeta}F\biggl(-\frac{i{\bar \zeta}}{2\sqrt{\eta }}\biggl)\biggl\},
\end{align} 
where complex functions $\eta $, $\zeta $, and ${\bar \zeta}$ are now
\begin{align}
\eta(t) = \frac{1}{4\sigma^2}-\frac{im}{2\hbar t},\quad \zeta(x,t) =\frac{x_0}{2\sigma^2}+i\frac{p_0}{\hbar}-\frac{imx}{\hbar t},\quad {\bar \zeta}(x,t)= \frac{x_0}{2\sigma^2}+i\frac{p_0}{\hbar}+\frac{imx}{\hbar t},
\end{align} 
while the time-dependent solution with the Neumann boundary condition is  
\begin{align}
\Phi(x,t)
=&\sqrt{\frac{m}{2\pi i\hbar t}}e^{-x_0^2/4\sigma^2+imx^2/2\hbar t} \nonumber \\
&\times \biggl[\frac{1}{2\eta}\biggl\{\frac{1}{2}\sqrt{\frac{\pi}{\eta }}\biggl(\zeta e^{\zeta^2/4\eta }+{\bar \zeta}e^{{\bar \zeta}^2/4\eta }\biggl)+i\zeta F\biggl(-\frac{i\zeta}{2\sqrt{\eta }}\biggl)+i{\bar \zeta} F\biggl(-\frac{i{\bar \zeta}}{2\sqrt{\eta }}\biggl)+2\biggl\} \nonumber \\
&-\biggl(\frac{x_0}{2\sigma^2}+i\frac{p_0}{\hbar}\biggl)^{-1} \biggl\{\frac12\sqrt{\frac{\pi}{\eta }}\biggl(e^{\zeta^2/4\eta }+e^{{\bar \zeta}^2/4\eta }\biggl)+\frac{i}{\sqrt{\eta }}F\biggl(-\frac{i\zeta}{2\sqrt{\eta }}\biggl)+\frac{i}{\sqrt{\eta }}F\biggl(-\frac{i{\bar \zeta}}{2\sqrt{\eta }}\biggl)\biggl\}\biggl].
\end{align}

\end{document}